\newcommand{\tsecompldate}{18th June 2004}
\newcommand{\tseprepno}{Imperial/TP/3-04/15} 
\newcommand{\tsehepphno}{\texttt{hep-ph/0406202}}

\documentclass[12pt,a4paper]{article}

\usepackage{graphics}

\usepackage{amssymb}
\usepackage{amsmath}

\usepackage{array}

\newcommand{\vol}[1]{\textbf{#1}}
\newcommand{\yr}[1]{(#1)}
\newcommand{\journal}[1]{#1}

\newcommand{\tarttitle}[1]{``#1'',}

\newcommand{\tinproctitle}[1]{``#1''}
\newcommand{\tbktitle}[1]{``#1''}

\newcommand{\tref}[1]{(\ref{#1})}

\newcommand{\tpre}[1]{}
\newcommand{\tprenote}[1]{}

\newcommand{\tnote}[1]{}
\newcommand{\tcomment}[1]{}
\newcommand{\tcommentx}[1]{}

\newcommand{\href}[2]{#2}
\newcommand{\eprint}[1]{\texttt{#1}}
\newcommand{\tseurl}[1]{\texttt{#1}}

\newcommand{\tsedevelop}[1]{{}}

\newcommand{\half}{\frac{1}{2}}
\newcommand{\bea}{\begin{eqnarray}}
\newcommand{\eea}{\end{eqnarray}}
\newcommand{\beq}{\begin{equation}}
\newcommand{\eeq}{\end{equation}}

\newcommand{\bi}{\begin{itemize}}
\newcommand{\ei}{\end{itemize}}

\newcommand{\ra}{\rightarrow}
\newcommand{\lra}{\longrightarrow}

\newcommand{\tsemat}[1]{{\mathbf{\textsf{#1}}}}

\newcommand{\unitmat}{\tsemat{\hbox{{\bf 1}\kern-.28em{\bf I}}}}
\newcommand{\dslash}{{d\kern-.45em{/}}}
\newcommand{\deltasl}{{\delta\kern-.28em{'}}}

\newcommand{\tsevec}[1]{{\boldsymbol{#1}}}

\newcommand{\chibar}{\bar{\chi}}
\newcommand{\phibar}{\bar{\phi}}

\newcommand{\pvec}{\tsevec{p}}

\newcommand{\jbar}{\bar{\jmath}}

\newcommand{\vecp}{\tsevec{p}}
\newcommand{\vecv}{\tsevec{v}}
\newcommand{\barvecv}{\bar{\tsevec{v}}}
\newcommand{\vvec}{\tsevec{v}}

\newcommand{\xvec}{\tsevec{x}}
\newcommand{\yvec}{\tsevec{y}}

\newcommand{\tseline}[1]{\mathsf{#1}}
\newcommand{\eline}{\tseline{e}}
\newcommand{\fline}{\tseline{f}}

\newcommand{\calD}{\mathcal{D}}
\newcommand{\calL}{\mathcal{L}}

\newcommand{\calZ}{\mathcal{Z}}

\newcommand{\bbR}{\mathbb{R}}

\newcommand{\SUL}{S_\mathrm{UL}}
\newcommand{\SNUL}{S_\mathrm{NUL}}

\newenvironment{acknowledgements}{\section*{Acknowledgements}}{ }

\newcommand{\expv}[1] {\ensuremath{\big\langle #1 \big\rangle}}
\newcommand{\oexpv}[1] {\expv{#1}_1}
\newcommand{\zexpv}[1] {\expv{#1}_0}
\newcommand{\cexpv}[1] {\expv{#1}_\mathrm{C}}

\newcommand{\diff}{\text{d}}
\newcommand{\Diff}{\mathcal{D}}
\newcommand{\tr} {\text{tr}}
\newcommand{\pderiv}[2]{\ensuremath{\frac{\partial #1}{\partial #2}}}
\newcommand{\pderivn}[3]{\ensuremath{\frac{\partial^{#3} #1}
    {\partial #2^{#3}}}}

\newcommand{\sumn}[1] {\sum_{#1 \in \Lambda}}
\newcommand{\prodn}[1] {\prod_{#1 \in \Lambda}}
\newcommand{\sumnn}[2] {\sum_{#1 \in \Lambda} \sum_{#2 \in
    \mathcal{N}_{#1}^+}}

\newcommand{\intall}{\int_{-\infty}^{+\infty}}
\newcommand{\intpos}{\int_0^{+\infty}}

\begin{document}

\typeout{--- Title page start ---}

\renewcommand{\thefootnote}{\fnsymbol{footnote}}

\begin{flushright}
\tseprepno \\
\tsehepphno \\
\tsecompldate \\
\tsedevelop{ (LaTeX-ed on \today ) \\}
\end{flushright}

\begin{center}
{\Large\bf Testing an Optimised Expansion on $Z_2$ Lattice Models}\\[1cm]
{\large T.S.\ Evans\footnote{E-mail: {\tt T.Evans@imperial.ac.uk},
WWW: \tseurl{http://theory.ic.ac.uk/$\sim$time/} }, M.\ Ivin }\\[1cm]
Theoretical Physics, Blackett Laboratory, Imperial College London,\\
Prince Consort Road, London SW7 2AZ  U.K.
\end{center}

{}\tnote{tnotes such as this not present in final version}

\renewcommand{\thefootnote}{\arabic{footnote}}
\setcounter{footnote}{0}

\begin{abstract}
We test an optimised hopping parameter expansion on various $Z_2$
lattice scalar field models: the Ising model, a spin-one model and
$\lambda \phi^4$. We do this by studying the critical indices for
a variety of optimisation criteria, in a range of dimensions and
with various trial actions.  We work up to seventh order, thus
going well beyond previous studies.  We demonstrate how to use
numerical methods to generate the high order diagrams and their
corresponding expressions.  These are then used to calculate
results numerically and, in the case of the Ising model, we obtain
some analytic results.  We highlight problems with several
optimisation schemes and show for the best scheme that the
critical exponents are consistent with mean field results to at
least 8 significant figures.  We conclude that in its present
form, such optimised lattice expansions do not seem to be
capturing the non-perturbative infra-red physics near the critical
points of scalar models.
\end{abstract}

\section{Introduction}

The aim of this paper is to study the LDE (Linear Delta
Expansion), a non-perturbative method. Finding alternatives to
weak-coupling perturbation expansions is vital as there are many
problems where such methods fail even when the theory is weakly
interacting, e.g.\ at phase transitions. Perhaps the most accurate
non-perturbative method is lattice Monte Carlo, but this has
trouble with fermions, non-zero charge densities and dynamical
problems \cite{MM}.\tnote{As our universe is expanding and for
most of its existence has had a conserved non-zero fermion
density, many problems can not be tackled using Monte Carlo
methods.} However, there are very few alternatives: large-N for
continuum calculations is one example, a hopping parameter
expansion on the lattice is another.  Unfortunately, these
non-perturbative methods can be poor e.g.\ see the expansions for
critical exponents $\eta$ and $\gamma$ in three dimensions using
large-N approximations in \cite{ZJ}.\tnote{See page 664
\cite{ZJ}.} Thus it is important to develop alternative
non-perturbative methods.

LDE is just such an alternative non-perturbative method.
Basically, it is an optimisation of an existing standard expansion
scheme. The earliest appearance of the term LDE we know is in
\cite{DM88,Jo}\tnote{According to \cite{BR02a}.} and some of this
emerged from the logarithmic Delta Expansion work of
\cite{Betal87,Betal88,BJ88a,BJ88b}. However LDE is such a generic
scheme that it has appeared in many contexts, has been
rediscovered many times and under many names: Optimised
Perturbation Theory \cite{St81,CH98}, Gaussian Effective Potential
approximation \cite{St84}, Variational Perturbation Theory
\cite{SSS,Kl93,JK95,KSF}, Order Dependent Mapping \cite{ZJ},
Screened Perturbation Theory \cite{KPP97}, method of self-similar
approximation \cite{Yu}, the variational cumulant expansion
\cite{WZZSDYX}, action-variational approach \cite{KM}. The method
has been applied to the evaluation of simple integrals
\cite{Jo,BDJ,BeDJ,Ok87,PO98}, solving non-linear differential
equations \cite{BMPS}, quantum mechanics
\cite{Jo,St81,St84,JK95,Ok87,PO98,Ca,Ki,DJ,GKS} and to quantum
field theory both in the continuum
\cite{St81,CH98,SSS,KPP97,Ok87,Ok87b,GM,OS98a,Chi00,Meu,BR02a,OS99}
and on a lattice (discussed below).\footnote{The wide
applicability of this approach under different guises means our
list in this paragraph is only a sample of LDE type papers,
suitable as a starting point for further research.  We do not
claim to establish the precise history of the method. Also see
\cite{BR02a} for comments about names.}

An important and recurring question is how good is LDE?  It
appears to be a quite arbitrary scheme though, as we will try to
highlight in the next section, many of the supposed limitations of
LDE are common to all other expansion methods. Likewise, widely
accepted methods are not always as good as one might have
supposed, see for example the large-N expansion of critical
exponents in \cite{ZJ}. LDE has, in fact, been tested rigourously
in QM where it has been shown to lead to fast convergence of
calculations of ``zero-dimensional'' path integrals
\cite{BDJ,BeDJ} and of the anharmonic oscillator
\cite{JK95,DJ,GKS}. However, such rigour is probably unobtainable
in QFT (see \cite{PO98} for an interesting discussion), where in
practice the only simple test is to compare results against
physical results, or, failing that, Monte Carlo calculations.

One of our goals was to look at the accuracy of LDE calculations
on the lattice, where we are in effect optimising hopping
parameter expansions. The method can go beyond Monte Carlo
results, e.g.\ one can easily handle finite charge densities
\cite{EJW}, but here the idea is to focus on calculations where
there are good Monte Carlo results, if not physical measurements,
as these can then provide us with the `true' answer against which
to test our results. We chose to look at critical exponents in
$\phi^4$ field theories as these are computationally accessible,
they have been studied extensively theoretically and they have
physical relevance, as shown by the available experimental data
for critical exponents \cite{KSF}.

Another goal was to extend the accuracy of existing LDE lattice
studies.  The literature on lattice LDE contains work on scalar
theories
\cite{Jo,WZZSDYX,AJ93a,AJ93b,AJP,ZTW,Ya,YWZ,ZL,TZ,EIM,Win}, pure
gauge models \cite{Jo,KM,AJ93a,AJ93b,AJP,ZTW,Ya,YWZ,ZL} and gauged
Higgs models \cite{EJR98a,EJR98b}.  Where they have made
comparisons with other techniques, the conclusions are usually
positive.   For instance (Jones strong coupling in gauge theory)
or the comparison of gauge Higgs models near transition
\cite{EJR98a,EJR98b}.  However, as far as we know, all the work in
the literature, except one pure gauge study \cite{KM}, has worked
only to third order in the expansion, doing the expansions by
hand.  To go to higher orders requires a computerised approach to
the expansion. Thus our second aim is to investigate how this
could be done.

The codes we developed were easily adapted for studies of both the
full $\phi^4$ QFT and spin models.  Thus we will also look at the
Ising and spin-1 models which fall into the same universality
class as the full $\phi^4$ model \cite{ZJ,KSF,HPV}. The
feasibility of LDE for full field theories at higher orders will
therefore be addressed while the numerical simplicity of the spin
models can be exploited at various points. Specific results will
be given for all three models.

In the next section we will outline the LDE in the context of our
$\phi^4$ field and spin theories on the lattice.  Section three
will examine our numerical results in detail, comparing these with
our more limited analytical results in section four.  The final
section we summarise our conclusions.

\section{The Free Energy Density in LDE}
\label{secforCtFED}

\subsection{General Discussion of LDE}

We will start by developing the LDE in very general terms by
considering a theory described by the full action $S$ which
depends on physical parameters $\vecp$. We will be looking at
scalar field theories where $\vecp$ will be masses and coupling
constants.

It is helpful to focus on the calculation of a specific quantity,
though the principles are the same for all quantities, so consider
the partition function $Z$
\begin{equation} \label{Zdef2}
Z(\vecp) := \tr \{ e^{- S}\} .
\eeq
The starting point for all expansions in QFT is that with the
physical action for the theory of interest, $S=S(\vecp)$, there is
no exact solution\footnote{In some rare cases in QFT, we may be
trying to construct an exact solution via some expansion, e.g.\ as
in two-dimensional sine-Gordon models.} for $Z$. We therefore turn
to some solvable theory described by an action $S_0=S_0(\vecv)$
which depends on a set of parameters $\vecv$.  These parameters
$\vecv$ of the solvable action $S_0$ will be specified later but
for now we just note they may be different from the physical
parameters $\vecp$ of the full action $S$.  The idea is to expand
about the trial action $S_0$, exploiting its solubility. To do
this we replace the action $S$ by $S_\delta$ the
$\delta$-\emph{modified action}, i.e.\
\begin{equation} \label{iSdelta}
S(\vecp) \longrightarrow S_\delta (\vecp, \vecv)= S_0 (\vecv)-
\delta \big( S_0 (\vecv) - S(\vecp) \big)
\end{equation}
Then rather than considering the complete partition function $Z$
of \tref{Zdef2}, we look at the \emph{$\delta$-modified partition
function} $Z_\delta$
\begin{eqnarray}
\label{iZdeltadef} Z(\vecp) = \tr \, e^{- S}
 &\lra&
 Z_\delta(\vecp, \vecv)
  = \tr\{  e^{- S_\delta(\vecv)} \}
  = \tr \{  e^{- S_0 (\vecv)} e^{-\delta \left( S_0 (\vecv) - S(\vecp) \right)} \}
\end{eqnarray}

Clearly we need to set $\delta=1$ at the end of the calculation to
return to the appropriate action of the physical theory,
$\lim_{\delta\ra 1} Z_\delta = Z$.  Thus the $\delta$ is
\emph{merely a book keeping parameter} introduced to keep track of
the terms in the expansion about $S_0$
\begin{subequations} \label{iZexpinfty}
\begin{align}
\label{iZexpinfty1} Z_\delta &
 = \sum_{n = 0}^\infty \frac {\delta^n} {n!} Z_n(\vecp,\vecv)
 \\
\label{iZexpinfty2} Z_n (\vecp,\vecv) &
 = \tr \left[ e^{- S_0 (\vecv)} \left( S_0 (\vecv) - S(\vecp) \right)^n \right]
\end{align}
\end{subequations}
where we have defined $Z_n$ as the $n^\mathrm{th}$ order term in
the expansion of $Z_\delta$. The expansion is in terms of
increasing powers of $S-S_0$ and the parameter $\delta$ is just a
parameter of technical use that records the order in the expansion
of any given term in the calculation.  Note that without further
information, demanding $|\delta| \ll 1$ is neither necessary nor
sufficient to control the convergence of this series, so we need
not be concerned that $\delta=1$ is required at the end of the
calculation.

Formally, summing all terms gives back the insoluble theory $S$
when $\delta=1$, but by assumption this can not be done here.
However, we are free to choose $S_0$ so that individual terms in
the expansion, $Z_n$, are obtainable, and in particular, the low
order terms can be calculated in a reasonable amount of time. This
is one major principle behind the choice of $S_0$. The idea is
then to truncate the series, i.e.\ we consider
\begin{equation} \label{iZexpR}
Z_\delta^{(R)}
  = \sum_{n = 0}^R \frac {\delta^n}{n!} Z_n
\end{equation}
where the superscript of $Z_\delta^{(R)}$ denotes the order at
which the expansion is truncated.

So far, all we have done is describe generic perturbation schemes.
For later comparison, let us consider standard (weak-coupling)
perturbation theory in this context.  Then $S-S_0$ would be
proportional to some small real coupling constant. For instance,
in the $\phi^4$ theories studied here we would choose $S-S_0=-
\int d^4x \lambda \phi^4$.  Then we hope that for $0\leq\lambda
\ll 1$ the small parameter $\lambda$ would ensure convergence, or
at least that a truncated series gives good answers.  In fact the
former is clearly not true, QFT perturbation series do not usually
converge, and the latter can also be false. The instability
expected for $\lambda$ negative, however small, indicates that
physical results will not be analytic in the complex $\lambda$
plane about $\lambda=0$. Thus a series in $\lambda$ can not be
convergent. The behaviour of such series is hard to study in full
QFT though it can still provide useful information by using ideas
such as Borel summability.\footnote{Rigourous results for LDE are
possible in one space-time dimension when the problem becomes one
of the Quantum Mechanics of an anharmonic oscillator.  See also
\cite{PO98} for a useful discussion of LDE and expansions in
general.} However, whether the first few terms are a good
approximation, despite these issues, changes depending on the
question being asked of the model. For instance, while
weak-coupling perturbation theory gives accurate results for QED,
at phase transitions weak-coupling perturbation theory is known to
break down even in weakly-interacting theories. Ultimately then,
all expansions in QFT are developed using physical intuition that
$(S-S_0)$ is suitably small, confirmed only by comparing with
physical results. It is only posthoc, and usually only in a
limited way  (if at all), that a rigourous mathematical basis for
the expansion can be given. QED is a successful example of this
process, and it shows that despite the mathematical uncertainties,
the procedure outlined so far is worth following. So here, as in
all expansions in QFT, a second principle guiding the choice of
$S_0$ should be that $S_0$ is chosen so that terms up to $n=R$ are
sufficient to give a good approximation.  Crudely speaking,
$(S-S_0)$ is to be thought of as ``small'', while the size of the
coefficient $\delta$ is one and does not control the convergence.

It is at this point that LDE departs from standard perturbative
expansions, and where the non-perturbative aspects come in.  In
the LDE approach the parameters $\vecv$ of the trial action $S_0$
are fixed using a variational method.  These \emph{variational
parameters} $\vecv$ lie at the heart of the LDE method and provide
the mechanism through which non-perturbative behaviour is
introduced into the model. Before the expansion is truncated, upon
substitution of $\delta = 1$, the dependence on the variational
parameters vanishes and we have the full theory which only depends
on the physical parameters $\vecp$ in $S$, so $Z=Z(\vecp)$.
However, a truncated series $Z_\delta^{(R)}$, will have residual
dependence on these unphysical variational parameters $\vecv$ even
when $\delta=1$, i.e.\
$Z_{\delta=1}^{(R)}=Z_{\delta=1}^{(R)}(\vecp,\vecv)$. The final
criteria for the choice of $S_0$ is that it has some variational
parameters $\vecv$ suitable for our subsequent exploitation.

Clearly the dependence in the truncated answer on the arbitrary
variational parameters $\vecv$ is unphysical. Thus the final stage
of the LDE method is to fix the variational parameters. Most
importantly, this is to be redone every time we repeat a
calculation at a higher order. Thus the values we assign to the
variational parameters depends on the order at which we truncate
the expansion so we are expanding about \emph{different} theories
$S_0$ at each order. Thus this procedure has been called an order
dependent mapping\tnote{See page 885 of \cite{ZJ}.} \cite{ZJ}. By
contrast, if $\vecv$ was to be fixed at some specific order of the
expansion and then used for all other orders, we would end up with
nothing else than another perturbative expansion.

The aim is to fix the variational parameters to values which will
produce a result closest to the true physical one. The problem in
achieving this goal is that there is no unique prescription which
tells us how to do this. We know of five broad categories of
methods used to fix the variational parameters.  PMS (the
principle of minimal sensitivity) and FAC (fastest apparent
convergence) are the two most common. The names PMS and FAC, and
their systematic definitions were introduced in \cite{St81}.
However, prior to this we find a PMS-like criterion used in a
study of a variational Hartree-type expansion applied to $\phi^4$
field theory \cite{Chang} and used in QM \cite{Ca,Ki}, and the FAC
criterion used in \cite{HS} to calculate the energy levels of an
anharmonic oscillator by a variational perturbative expansion. In
continuum calculations, the use of gap equations (choosing
variational parameters such that the self energy is zero and the
full propagator has a pole at the variational mass) is a third
approach.  A fourth is that of Meurice \cite{Meu} who cut off the
range of integration of the fields.

However we shall focus largely on a fifth approach and fix our
variational parameters by demanding that they minimise the free
energy \cite{WZZSDYX,EIM}. This choice is simply based on the
physical principle that the free energy density $f(\vecp)$ of a
system always tends to a minimum. Thus we seek a minimum of the
function $f^{(R)} (\vecp, \vecv)$ in the variational parameter
phase space $\vecv$ whilst keeping the physical parameters $\vecp$
constant during the optimisation. This will fix the variational
parameters to the optimum value, which we will denote with a bar
as $\bar {\vecv}$ --- a function of the physical parameters
$\vecp$ and the order $R$. Likewise the optimum free energy
density will then be
\beq
\bar{f}^{(R)} (\vecp):=f^{(R)} (\vecp, \bar{\vecv}) \leq
 f^{(R)} (\vecp, {\vecv}) \;\;\; \forall \; \vecv ,
\eeq
for a given set of physical parameters $\vecp$. Assuming
analyticity in $\vecv$ near $\bar{\vecv}$, we have
\begin{equation} \label{iPMSdef}
\pderiv {} {\vecv} f_\delta^{(R)} = 0
\end{equation}
At this optimal point $\vecv = \bar{\vecv}$, `small' variations in
the components of $\vecv$ produce `negligible' variations in
$f_\delta^{(R)}$, thus we are as close to being independent of the
variational parameters as we will ever be.  We will compare this
minimum free energy principle against the other methods later in
section \ref{sisopt}.

\subsection{Lattice scalar field models and LDE}\label{slsfm}

Now let us apply this to specific models. We will consider
theories of a single scalar field on a hyper-cubic Euclidean
space-time lattice\footnote{See conclusions in section \ref{scon}
for comments on hyper-tetrahedral lattices.} with lattice spacing
set to be $1$ and of various dimensions with action
\begin{eqnarray}
S &=& \SUL+ \SNUL
   \label{Slatdef}
\\
\SUL &=& \sum_{i \in \Lambda}
  J \phi_i +
  \alpha \phi_i^2 +
  g \phi_i^4
    \label{SULdef}
 \\
 \SNUL &=&
 - \kappa \sum_{i \in \Lambda} \sum_{j
\in \mathcal{N}_i^+}
  \phi_i \phi_j
    \label{SNULdef}
\end{eqnarray}
We have split the action into two types of term. $\SUL$ contains
the ultra-local terms, involving products of fields from the same
lattice point.  The non ultra-local terms $\SNUL$ have products of
fields at different lattice points. The generic physical
parameters $\vecp$ of the full action $S$ are $\vecp = (\kappa, J,
\alpha, g)$, though one can rescale the fields and remove any one
of these parameters. $\Lambda$ is the set of space-time points,
$\phi_i$ is the field value at site $i$, and $\mathcal{N}_i^+$ is
the set of nearest neighbours to lattice point $i$ in a positive
direction.

For a full field theory we choose $\phi_i \in \bbR$ and exploit
the freedom to rescale the field to set $\kappa=1$ so the physical
parameters are then $\vecp_\mathrm{qft}=(J, \alpha, g)$. To make
contact with the the usual notation used for the spin-1 model, we
set $g = 0$, use the rescaling freedom to choose $\alpha=1$ so
then $\vecp_\mathrm{spin 1}=(\kappa, \alpha)$ and restrict
$\phi_i$ to $\phi_i \in \{ +1, 0, -1 \}$.  For Ising models we
further restrict $\phi_i \in \{ +1, -1 \}$ but we still have
$\vecp_\mathrm{ising}=(\kappa)$. All these models belong to the
same universality class as the lattice $\phi^4$ model
\cite{ZJ,KSF,HPV}.

The first step in LDE is to chose the trial action $S_0$ according
to the three principles set out above, namely that $S_0$ describes
a theory that we can solve, $S_0$ is a reasonable approximation to
the full theory, and $S_0$ contains non-physical, variational
parameters.

On a lattice, an action consisting of only ultra-local terms is
soluble since the path integral factorises because the integral
over the field at each space-time point is then independent of all
the other field integrals. The insolubility of the full lattice
$\phi^4$ action \tref{Slatdef} comes from the non-ultra-local
terms $\SNUL$ in the kinetic terms \cite{MM}. Thus we must choose
a pure ultra-local action for our trial action $S_0$. Bearing in
mind our second and third criteria for choosing a good $S_0$, a
suitable guess for $S_0$ would be a trial action of the same form
as the ultra-local part of the full action $S$ but with arbitrary
coefficients.  Thus we introduce
\begin{equation} \label{S0latdef}
S_0 (\vecv) := \sum_{i \in \Lambda} L_0(\phi_i,\vecv) ,
\end{equation}
where our trial action is ultra-local (solvable) and depends on
the variational coefficients $\vecv$.  One choice is to make this
of the same form as the ultra-local terms of the full Lagrangian,
\tref{SULdef}, to be a close approximation to the full theory.  So
we will work with
\begin{equation} \label{L0latdef}
 L_{0} (\phi,\vecv) := \left[ j \phi + k \phi^2 + l  \phi^4 \right]
 \;\;\;\;\;\;\;
 \vecv:=( j, k, l)
\end{equation}
The variational parameters, $j$, $k$ and $l$, are to be fixed by
some optimisation condition once the expansion has been performed.
In practice we do not always use the quartic variational parameter
and in the Ising model both quadratic and quartic terms are
trivial.  In the end we will be left with expressions in terms of
\emph{statistical averages} $\zexpv{\phi^p}$ taken with respect to
\tref{S0latdef} where
\begin{eqnarray}
\zexpv{\phi^p} (\vecv) &:=& \frac{1}{\calZ_0}
 \int \diff \phi \, \phi^p \exp\{- S_0(\vecv)\}
 = \frac{I_p}{I_0}
  \label{zexpdef}
\\
 \calZ_0 (\vecv) &:=& \int d \phi \, \exp\{- L_0(\phi,\vecv)\} = I_0
 \\
 I_p(\vecv) &:=& \int d\phi \; (\phi)^p \exp\{L_0(\phi,\vecv)\}  .
 \label{Ipdef}
\end{eqnarray}
Note that the ultra-local property of $L_0$ is crucial and allows
the factorisation into integrals over field values at a single
site.

This is a good place to note that the expansions we are interested
in are traditional lattice expansions, and it is the variational
aspect which is novel. If we were to fix our variational
parameters equal to the appropriate physical ones, i.e.\ $j = J$,
$k = \alpha$ and $l = g$ in \tref{S0latdef}, then we would have an
expansion in the the non-ultra-local term of $S$ only. This is
then a traditional strong coupling or \emph{hopping parameter}
expansion \cite{MM}. It is then not surprising that we can adapt
the machinery developed for such traditional expansions
\cite{MM,Wo,Englert63,Rushbrooke64,RBW,HR99,CHPPV} for the general
LDE problem.

\subsubsection{Trick for ultra local terms in $S-S_0$}

If we recall the definition of the $\delta$-modified action given
in \tref{iSdelta}, we see that we have to expand $\exp
\{\delta(S_0 - S)\}$ to a given order and evaluate the path
integral with respect to the ultra-local $\exp \{S_0 \}$. However,
parts of $(S_0 - S)$ are also ultra-local and no more difficult to
deal with than $S_0$.  We can exploit this and make the
diagrammatic evaluation below a great deal simpler.  Thus let us
make a further split of the action, replacing the physical action
$S$ by $S_{\delta\delta}$
\bea
S \longrightarrow S_{\delta\delta} &=& S_1 -  \delta_1 \SNUL,
  \label{ddeltadef}
 \\
 S_1 &:=& S_0 + \delta_2 (\SUL-S_0) = \sumn{i} L_{1}[\phi_i]
 \label{S1def}
  \\
 L_{1}[\phi_i]  &:=&
 (j + \delta_2(J-j)) \phi_i
 + (k + \delta_2(\alpha-k))  \phi_i^2
 + (l +\delta_2( g-l)) \phi_i^4
\label{L1def}
\eea
The idea is that we can do an expansion in $\delta_1$ first up to
the desired order R, treating $S_1 = S_0- \delta_2 (\SUL-S_0)$ as
an \emph{intermediate variational action}. The parameter
$\delta_2$ then counts the powers of $-\SNUL$ which contains all
the non local terms in the calculation.  As $S_1$ is ultra local,
this expansion in $\delta_2$ is just as feasible yet the
diagrammatic methods used below are much simpler since there is
only one type of term in $-\SNUL$.  This leaves us with an
expression for a quantity of interest, such as $Z$, of the form
\bea
Z_{\delta 1}^{(R)}
  &=& \sum_{n = 0}^R \frac {\delta_2^n}{n!} Z_{1n}
  \label{iZ1expR}
\\
 Z_{1n} (\vecv, \delta_1) & =& \tr \left\{ e^{- S_1 (\vecp,\vecv, \delta_1)}
 \left( S_0 (\vecv) - S(\vecp) \right)^n \right\}
 \label{Z1ndef}
\eea
We can see that to return to the proper LDE expansion of
\tref{iSdelta}, we want to set $\delta=\delta_1=\delta_2$ in
\tref{ddeltadef} and truncate the series rigourously to order $R$
in $\delta$. Thus we have to do a further expansion to order $R-n$
in $\delta_1$ of each term $Z_{1n} (\vecv)$.  In general this
would be doubling the work we have to do to produce an overall
$\delta$ expansion. However in this case we are expanding in
powers of $(S_0-\SUL)$ about $S_0$ and both are ultra local.  This
leads to this $\delta_1$ expansion being significantly simpler
than the first expansion in $\delta_2$ and the non ultra local
terms.  We will see this means that there is no need to use the
full diagrammatic machinery a second time. This will be clearer
when we have finished outlining the complicated $\delta_2$
expansion to which we now turn.

\subsubsection{Partition function expansion}

The expansions for all quantities can be expressed in terms of
what we call \emph{statistical averages}\tnote{Marko used
\emph{zero expectation value}, or \emph{zero average}. Why are
these not statistical averages?} $\oexpv{Q}$,  given in terms of
our intermediate variational action \tref{S1def}:
\begin{eqnarray}
\oexpv{Q} &:=&  \frac {1} {Z_1} \int \Diff \phi \, Q \, e^{- S_1}
 \label{oexpdef}
\\
  Z_1 &:=& \int \Diff \phi \, e^{-S_1}
 \label{Z1def}
\end{eqnarray}
For instance partition function can be written neatly  as
\begin{equation}
Z_{\delta 1} = Z_1 \sum_{n=0}^\infty \frac{\delta_2^n}{n!}
 \oexpv{(-\SNUL)^n }
 .
 \label{Zd1b}
\end{equation}

It is useful to introduce a diagrammatic notation at this point
and to use the language of graph theory (e.g.\ see
\cite{Wo,Bollobas}). The whole space-time lattice is a
\emph{graph} whose vertices are the space-time points.  The edges
of the lattice graph should be chosen to be those linking
neighbouring space-time points which appear in non ultra-local
products in the lattice action coming from the derivative terms.
For our simple nearest neighbour approximation of the derivatives,
the edges connect all nearest neighbour vertices.  Thus each edge
represents a unique nearest neighbour vertex pair and appears only
once in the edge set of the lattice graph, i.e.\ our lattice graph
is a \emph{simple graph}. The non ultra local action can then be
written as
\begin{eqnarray} \label{DSDiagDef}
-\SNUL &=& \sumnn{i}{j}
 \setlength{\unitlength}{1pt}
 \begin{picture}(20,10)
 \put(0,2){\circle*{4}}
 \put(0,2){\line(1,0){20}}
 \put(20,2){\circle*{4}}
 \put(0,6){$i$}
 \put(18,6){$j$}
 \end{picture}
 \\
\label{LineDef}
 \setlength{\unitlength}{1pt}
 \begin{picture}(20,10)
 \put(0,2){\circle*{4}}
 \put(0,2){\line(1,0){20}}
 \put(20,2){\circle*{4}}
 \put(0,6){$i$}
 \put(18,6){$j$}
 \end{picture}
  & :=& \kappa \phi_i \phi_j
\end{eqnarray}
From the diagrammatic point of view, performing the sum over $i$
and $j$ in $\SNUL$ is equivalent to running over all the elements
of the \emph{edge set} of the lattice graph.

What this means is that the $n$-th order term in the $\delta_2$
expansion of the partition function \tref{Z1def} is represented by
a sum over all possible ways of choosing $n$ edges from the edge
set. We will call each choice a \emph{configuration}.  However
note that we must include terms where we choose the same edge more
than once. Thus in the language of graphs, our configurations
include both simple and non-simple graphs.  Only the simple
configurations (i.e.\ ones which are simple graphs) are subgraphs
of the lattice graph and these will be our main focus.  We will
see that the expressions represented by non-simple graphs
represent straight forward generalisations of the expressions
represented by some simple graph. We will use $C$ to indicate a
configuration of the lattice.

Statistical averages have one important property given that we
have chosen our variational actions ($S_1$ here, $S_0$ ultimately)
to be ultra-local, namely the statistical averages of any
polynomial of fields is also factorisable into terms depending on
fields at only one lattice point. First the normalisation in our
statistical averages factorises
\begin{eqnarray}
Z_1 &=& \int \Diff \phi \, e^{- S_1} = \prodn{i} \int \diff \phi_i
\exp \left\{ - L_1(\phi_i) \right\}
 = (\calZ_1)^N,
  \label{Z1defb}
 \\
 \calZ_1 &:=& \int \diff \phi \exp \left\{ - L_1(\phi) \right\}
 \label{calZ1def}
\end{eqnarray}
where $L_1$ is defined in \tref{L1def}. Note that for simplicity
we are assuming here that the physical sources, $\vecp$, are
space-time constants, so that we obtain the same normalisation
factor $\calZ_1$ whatever lattice site $i$ is used in \tref{Z1def}
(translational invariance).

We can now write our statistical averages as\footnote{The
statistical averages $\zexpv{\phi^p}$ defined in \tref{zexpdef}
with respect to $L_0$ of \tref{S0latdef}have identical
factorisation properties but now in terms of the $I_p$ integrals
of \tref{Ipdef}.}
\bea
\oexpv{(\phi_1)^{n_1} (\phi_2)^{n_2} \ldots (\phi_i)^{n_i} \ldots
\ldots}
 &=&
  \frac{1}{Z_1} \int \Diff \phi \, e^{- S_1} (\phi_1)^{n_1} (\phi_2)^{n_2}
   \ldots  (\phi_i)^{n_i} \ldots \ldots
   \\
&=& \oexpv{(\phi_1)^{n_1}} \oexpv{(\phi_2)^{n_2}} \ldots
\oexpv{(\phi_i)^{n_i}} \ldots \ldots
 \label{1fact}
 \\
  &=& \frac{J_{n_1}}{J_0} \frac{J_{n_2}}{J_0}  \ldots \frac{J_{n_i}}{J_0}  \ldots \ldots
  \label{Statfact}
\eea
and they depend only on the integrals over fields at a single
space-time point
\bea
 J_{n}(\vecv,\delta_1) &:=&
 \int \ d\phi
  (\phi)^{n} e^{-L_1(\vecv,\delta_1,\phi)}
  \label{Jndef}
\eea
This gives us a simple translation from the statistical averages
associated with configurations $C$ to algebraic expressions:
\bea
\oexpv{C} &=& \kappa^n \prod_{i \in C_\mathrm{vertex}}
\frac{J_{k_i}}{J_0}
\eea
where $C_\mathrm{vertex}$ is the set of vertices in configuration
$C$.  $k_i$ is the connectivity of the i-th vertex, that is the
number of edges in the configuration $C$ which have one end at the
$i$-th vertex. The configuration has $n$ edges.

It is important to note that since we have translational
invariance, the expressions depend only on the connectivities of a
given configuration and not on the precise position of the
configuration on the lattice.  It is therefore useful to consider
graphs which are not embedded on the space-time lattice, and these
we call \emph{diagrams}, D.  Every diagram is isomorphic to many
configurations yet all these configurations represent the same
algebraic expression. The number of different ways we can embed
the diagram in the space-time lattice, that is the number of
configurations isomorphic to a diagram is called the \emph{lattice
constant} of the diagram.\tnote{Thus, in the diagrammatic
approach, we will be interested in counting the number of
different configurations (on the lattice) that give rise to the
same diagram. This number will be called the \emph{multiplicity}
of the diagram.} For instance the expression for the partition
function is then
\bea
Z_{\delta 1} &=& Z_1 \sum_{D}
 (\kappa \delta_2)^n \frac{b_D}{n!} c_D
 \left(\prod_{i \in D_\mathrm{vertex}} \frac{J_{k_i}}{J_0}\right)
 \label{Zd1c}
\eea
Here the sum is over all possible graphs, simple and non-simple,
connected and disconnected, as these are the set of diagrams $D$.
The $c_D$ is the lattice constant for the graph $D$ while $n$ is
the number of edges in the diagram. The last factor, $b_D$, is a
binomial factor coming from the fact that the same edge in a
configuration of $n$ edges, can come from any of the $n$ factors
of $\SNUL$ in the $(-\SNUL)^n$ terms.  Allowing for non simple
diagrams, we see that
\beq
\frac{b_D}{n!} = \prod_e \frac{1}{n_e!}
\eeq
where the product is over all \emph{distinct} edges, $e$,  in the
diagram, and $n_e$ is the number of times each edge appears in the
edge set of the diagram.\tnote{Marko's \emph{multiplicity} $m_D$
is then $m_D=b_Dc_D$, at least for simple diagrams} Thus for any
simple diagram, $b_D$ is always $n!$, while for a maximally
non-simple diagram, i.e.\ diagram of $n$ identical edges, $b_D
=1$.


\subsubsection{Free energy expansion}

The partition function is not the easiest quantity to calculate.
It involves all powers of the volume, here proportional to $N$ the
number of lattice sites.  The free energy density, $f$, should be
intensive and simpler to calculate.  However, as the logarithm of
the partition function it has a more complicated expression in
terms of statistical averages so it is convenient to work with
intermediate quantities called \emph{cumulant averages} (also
called \emph{cumulant expectation values} or
\emph{semi-invariants}) and the expansion in terms of these
averages is sometimes called a \emph{cumulant expansion} or a
\emph{cluster expansion}\footnote{The cumulant expansion owes its
origins to studies in chemistry in the late 1930's. Starting in
1959, a series of papers \cite{Brout1,Brout2,Brout3} described the
application of the expansion to the Ising model. Following papers
\cite{HC,SCH} further systematised the expansion, and applied it
to the Ising and Heisenberg models. More recently, we find the
cumulant expansion introduced and used in various ways and applied
to a wide range of models. Some examples are
\cite{KM,TZ,EIM,WZZSDYX}. The approach we take to presenting the
cumulant expansion is motivated by \cite{Wo}.}. We can define the
cumulant averages to play the same role for the free energy as the
statistical ones did for the partition function \cite{Englert63}.
Adapting to our case, where we are keeping $\delta_1 (\SUL - S_0)$
in the exponential and expanding initially just in powers of
$-\delta_2\SNUL$ and truncating at order $R$  we have by analogy
with \tref{Zd1b}
\begin{eqnarray}
 f := - \frac{1}{N}\ln(Z) , &&
 f^{(R)} := \sum_{n = 0}^R \frac{\delta_2^n}{n!} f_{1n}
 \label{fRdef1}
 \\
 f_{10}  = - \frac{1}{N} \ln Z_1 = -  \ln \mathcal Z_1
 , &&
 f_{1n} = - \frac{1}{N}\cexpv {\left( -\SNUL \right)^n}
 \label{fRdef2}
\end{eqnarray}
Note that we can use this to define the cumulant averages
$\cexpv{\ldots}$.

These cumulant averages have several important properties which we
will use to calculate them. First they are linear so that we can
use the diagrammatic notation
\beq
\cexpv {\left( -\SNUL \right)^n} = \sum_{D} b_{D} c_{D} \cexpv{D}
 \label{cumlin}
\eeq
where the sum is over all distinct order $n$ diagrams, that is
diagrams with n edges, and $b_D$ and $c_D$ are the binomial and
lattice constant factors for diagram D exactly as we had
above.\tnote{Is this true? Any extra factors here?}  Linearity
together with the translation invariance in our model ($\vecp$ and
$\vecv$ are space-time constants in our calculations) also means
that every diagram is repeated once for each space-time point.
Thus we can take advantage of this fact in future by calculating
the embedding factors $c_D$ of the graphs \emph{per lattice site},
at the same time discarding the factor of $\frac{1}{N}$ sitting in
front of the non-zero orders of the expansion in the above
equation. However it is important to note that unlike lattice
Monte Carlo calculations, our method works on a lattice of
infinite size.

A second property is the systematic link between cumulant and
statistical averages \cite{KM,Rushbrooke64,RBW}:
\begin{equation} \label{GenCumDef}
\cexpv{\Theta_{e_1} \cdots \Theta_{e_n}}
 = \sum_{\text{partitions
  of} \ (e_1 \ldots e_n)}
  (-1)^{k - 1} (k - 1)!
  \underbrace {\oexpv
  {\Theta_{e_1} \cdots \Theta_{e_r}} \cdots \oexpv {\Theta_{e_s} \cdots
  \Theta_{e_n}}}_{k \ \text{factors}}
\end{equation}
where $\Theta_e$ \emph{must} always be one of the operators whose
\emph{sum} makes $\SNUL$ in \tref{cumlin}. Thus here it is an
operator $\phi_i\phi_j$ associated with one edge $e=(i,j)$. It is
vital to note that cumulant averages do \emph{not} have a
factorisation property analogous to that of the statistical
expectation values \tref{1fact}.

Finally, we come to the crucially important \emph{cluster
property} of the cumulant averages. Given the ultralocal $S_1$ in
the exponentials, the statistical averages in \tref{GenCumDef}
factorise if \emph{any} of the operators $\Theta_{e_i}$ does not
depend on field values in any of the other $\Theta_e$ operators.
One can quickly check that in this case the cumulant average
vanishes. Put another way, the cumulant average of any
disconnected graph is always equal to zero. A simple proof by
induction can be found in \cite{RBW}, while the first general
mathematical proof was presented in \cite{Sherman64}.

This last property is of great importance to us, because it vastly
simplifies our task of calculating the $\cexpv {\left( \SNUL
\right)^n}$ terms.

If we put all of this together we arrive at a final formula
\begin{eqnarray}
 f^{(R)} &:=& \sum_{n = 0}^R \frac{1}{n!} f_{1n}
 \label{fRdef1b}
 \\
 f_{10}  = - \frac{1}{N} \ln Z_1 = -  \left[ \ln \mathcal Z_1 \right]_{R}
 , &&
 f_{1n} = - \sum_{D \in \calD_c^n}  b_{D} \frac{c_{D}}{N}
 \left[\cexpv{D}\right]_{R-n}
\end{eqnarray}
The sum is over the set $\calD_c^n$ of connected diagrams of
$n$-links.  The notation $[Q]_{m}$ indicates that we have to
expand the quantity $Q$ to order $m$ in a $\delta_1$ expansion
\beq
 \left[Q \right]_m := \sum_{j=0}^m \frac{1}{j!}
 \left.
 \frac{\partial^j Q }{\partial  (\delta_1)^j}\right|_{\delta_1=0}
\eeq
A worked example is provided in appendix \ref{secnumFED}.

\subsubsection{Numerical evaluation of free energy}

The first step is to find the lattice constants $c_D$.  To do this
we generate a complete set of \emph{simple} \emph{connected}
\emph{configurations} of up to $R$ edges, modulo translation
invariance.  The algorithm used was inspired by one used in
percolation theory \cite{SA,Me90} and is outlined in appendix
\ref{aguc}.

We then have to see which of these configurations are isomorphic
to a given diagram diagram $D$ so that we can obtain the lattice
embedding constant $c_D$ for each diagram.  To do this we use a
\emph{canonical labelling} for our configurations, that is for
each configuration we map it to an abstract graph that has a
unique labelling.  In this way we know that if and only if two
configurations are assigned to the same abstract graph are they
identical.  To do this we used a small part of McKay's beautiful
set of graph routines called \emph{nauty}\footnote{The name arises
by forming an acronym from \emph{No AUTomorphisms, Yes?}.}
\cite{nauty}.\tnote{The \emph{nauty} routines are implemented in
the \texttt{C} programming language, which enabled us to interface
with them our \texttt{C++} program generating the configurations.}

It is a simple matter of combinatorics to generate the $b_D$ and
$c_D/N$ for all the non-simple but connected diagrams $D$ up to
order $R$ by duplicating existing links in lower order simple
diagrams.\footnote{In our implementation, some of our non-simple
diagrams were isomorphic to others in our list but this redundancy
is relatively small.}

Thus this stage of our computation provided a list of all
connected diagrams up to order $R$, together with their $b_D
c_D/N$ coefficients for a specific lattice and in specific number
of dimensions. However, the list can be used for any model where
the only non-ultra local term is a nearest neighbour interaction.
Our compiled programme producing this list of diagrams was
relatively fast on, and used little memory of, a common desktop
computer, typically taking less than a day.\tnote{Example, times,
sizes?}

The next stages were implemented using the interpreted algebraic
manipulation programme, MAPLE \cite{Maple}.  This performed the
transformation of the cumulant averages represented by the
diagrams calculated previously $\cexpv{D}$, into the expressions
in terms of statistical averages, $\oexpv{\ldots}$.  It then had
to expand each statistical average in terms of $\delta_1$.  This
involves replacing each $J_p$ integral of \tref{Jndef} with the
appropriate expression in terms of $I_p$ integrals of \tref{Ipdef}
using the $L_0$ defined in \tref{L0latdef}. Moving from $J_p$ to
$I_p$ integrals requires explicit knowledge of the ultra-local
terms in both the action and the variational action and so is
highly model dependent. The complete expression for each diagram
$D$ is then truncated to order $R$ in $\delta=\delta_1=\delta_2$.
Putting the expressions for each diagram together with their
coefficients gives an expression for the free energy.  MAPLE then
outputs optimised $C$ language routines for the free energy
expressions. Thus each run providing a routine for a specific
model in a specific number of dimensions and on a specific type of
lattice. Results were checked by taking appropriate limits (easy
in MAPLE) of code for the third order expressions produced
independently in an earlier study \cite{EIM}.\tnote{In the same
manner, the same MAPLE programme also provided $C$ routines for
derivatives of the free energy, such as the expectation value of
$\phi$ and its susceptibility.}

The advantages of using MAPLE were that development is relatively
easy, it is extremely easy to alter the code to suit different
models or to produce code for derivatives of the free energy, and
it produces fast $C$ routines for the final stage of our analysis.
The disadvantage is that as interpreted code it is relatively
slow. Never the less, using only basic optimisations, the MAPLE
code for seventh-order ran just within the memory (1Gb) and speed
(one day) limitations of a desktop computer.  It is clear that to
go to higher orders a compiled version of these algorithms would
have to be used. The main job of the MAPLE code is to produce and
manipulate truncated polynomials with a relatively limited types
of coefficient.  Thus writing explicit compiled code for this part
is feasible.

The third and final stage of our numerical evaluation was
implemented as another compiled programme.  Using the free energy
routines provided by the MAPLE programme, the task was to find a
set of values for the variational parameters $\vvec$ which
minimised the free energy for a given set of physical parameters
$\pvec$. The $I_p$ integrals required (for the full field theory
case) are relatively straightforward to integrate, essentially
dominated by a single peak in the integrand, so standard routines
of \cite{NR} are sufficient.  Our experience also showed that
standard techniques to find the minimum in the multi-dimensional
variational parameter space \cite{NR}, that defined by $\vecv$,
were successful. The practical problems encountered lay elsewhere.

The first problem is that the expressions for the full seventh
order free energy are enormous.  The files for the free energy
$f^{(R)}$ grew as follows: $R=1$ used  6Kb, $R=3$ used 40Kb, order
$R=5$ used 0.5Gb and for $R=7$ we had a 5Gb file.  We found we
needed about $1.5Gb$ of memory to compile our seventh order
routines. A rough guess for the size of the 9th order code file
would be approximately 50 gigabytes in size, beyond the limits of
desktop computing.

However, this size indicates part of another serious problem,
namely numerical accuracy.  Our 7th order code is a sum of a vast
number\tnote{How many?} of terms, each term being a product of
several variational and physical parameters, some very large
integer constants $b_D c_d/N$, and several $I_p$ integrals.  Each
of the $I_p$ integrals must in turn be evaluated numerically
involving many additions.  Worse near a critical point where
typically $j \ra 0$ there is a large variation in the size of the
terms.  Some, such as $I_p$ for $p$ odd, approaching zero as $j
\rightarrow 0$. Further, we are interested in variations of the
free energy as its parameters are altered by small amounts.  For
instance, to locate the minima in variational parameter space, we
will often need to compare $f^{(R)}(j=0)$ against
$f^{(R)}(j=\epsilon)$ for small $\epsilon$. Variations in the
physical parameters are used to calculate physical quantities such
as $\expv {\phi}$. Thus tiny terms in the expressions for the free
energy may well be significant for the overall change in the free
energy.  Just as bad, it is important to ensure that the
variations in large terms, which may be only small fractions of
the total value of that term, are also accurately calculated. A
careful study of even the third order calculations suggests that
the double precision accuracy available on typical \verb$C++$
compilers for desktop PCs may not be sufficient. Thus we
implemented our codes using arbitrary precision arithmetic,
provided by the GNU multiple precision routines \cite{GMP}. This
enabled us to choose the accuracy of all our computations at the
start of a run.  For instance we were able to use 256 bit
arithmetic providing about 70 decimal places of accuracy.  Further
details and examples are provided in appendices \ref{secnumNA} and
\ref{abhpa}.

\subsection{Physical Quantities} \label{secnumPQ}

The free energy is not the only quantity of interest. We can also
calculate various derivatives of the free energy and it is the
behaviour of these at the critical point which are directly linked
to critical exponents.

Let us use the example of $\expv \phi$ to illustrate issues which
apply to all our calculations of derivatives of the free energy.
Not only is this quantity the order parameter for the expected
phase transition, but it is also the quantity which will give us
an estimate of the critical exponent $\beta$ (see equation
\eqref{betadef}). Analytically, $\expv {\phi}$ is related to the
free energy through
\begin{equation} \label{dfdJ}
\expv {\phi} = \pderiv {f} {J}
\end{equation}
However, we are not dealing with the full free energy density, but
rather with the truncated version $f^{(R)}$. For this reason, we
shall take a closer look at how we can make use of the above
equation to calculate the field expectation value.

One way of proceeding is to use the values of our optimised free
energy $\bar{f}^{(R)} (\vecp) := {f}^{(R)} (\vecp,\barvecv)$
directly to numerically estimate $\expv {\phi}$. After all, once
we have optimised $f^{(R)}$, we do have our best guess for the
free energy and that is meant to contain all the thermodynamic
information. The idea is straightforward to implement using
\begin{equation}
\expv {\phibar}^{(R)}_\mathrm{diff} \approx \frac {\bar{f}^{(R)}
(J + \epsilon) -
      \bar{f}^{(R)} (J)} {\epsilon}
      \label{phidiffdef}
\end{equation}
This way, we have to make two optimisations of $f^{(R)}$: once
using some value $J$ for the physical source, and another time
using $J + \epsilon$, where $\epsilon$ is some very small number.
Note that this involves taking the difference of two free energies
which may be very similar in size and we are only able to do this
because we can choose whatever accuracy we require for our
numerical calculations.

The alternative approach is to produce explicit code for $\expv
{\phi}^{(R)}(\vecp,\vecv)$ and to set the variational parameters,
$\barvecv=\barvecv(\vecp)$ to give a result
\beq
\expv
 {\phibar}^{(R)}_\mathrm{expl} (\vecp)  =
 \expv {\phibar}^{(R)}_\mathrm{expl} (\vecp,\barvecv).
 \label{phiexpldef}
\eeq
The subscripts on the ${\phibar}^{(R)}$ indicate where the result
was obtained via differentiation of optimised free energies (diff)
or via substitution of values of $\vecv$ which optimise the free
free energy into the explicit expressions for ${\phibar}^{(R)}$
(expl).  This is straightforward in principle as it is relatively
trivial to implement the derivatives required by \tref{dfdJ} in
MAPLE on the general algebraic expressions, and then it can
produce optimised \verb$C$ routines for the final numerical
evaluation, just as we did for the free energy.

The difference is with ${\phibar}^{(R)}_\mathrm{diff}$ we must
produce optimised variational parameters $\bar{\vecv}(\vecp)$,
functions of the physical parameters $\vecp$, at two physical
values, $J+\epsilon$ and $J$ while for
${\phibar}^{(R)}_\mathrm{exp}$ we only optimise the free energy at
$J$ (other physical parameters held constant).

However, it is not quite so simple and it is instructive to have a
more detailed look at this explicit code approach to the physical
quantities. From equation \eqref{dfdJ} we see that
\begin{equation} \label{dfdJ1}
\expv {\phi} = \pderiv {f} {J} = - \frac{1}{N} \frac{1}{Z}
\pderiv{Z}{J}
\end{equation}
where the generating function $Z$ is of the form
\begin{equation} \label{Zexplicit}
Z = \tr \ e^{- S_0 (j, k, l)} e^{\delta \Delta S (J,
  \alpha, g; j, k, l)}
\end{equation}
with $\Delta S= S_0-S$. The second exponential in the above
equation is the one that is expanded and truncated at some order
$R$. Also in this term is the $\Delta S$ term that carries all the
$J$ dependence. Thus, by using equation \eqref{dfdJ1} as a basis
for calculating $\expv \phi$, i.e.\ differentiating $f$ by the
physical source $J$, we would bring down a power of $\delta$ as
well as a factor of $\phi$.  If this was done on expressions for
$f^{R}$, then we would actually be producing an expression for
$\delta \expv {\phi}$ which has only $(R-1)$ terms in the delta
expansion, and we have $\expv {\phi}$ only up to
$O(\delta^{R-1})$. Put another way, the lowest order term in
$f^{(R)}$ is $\ln(Z_0)$ which is independent of physical
parameters like $J$ and this the term lost.

One solution is to develop a new diagrammatic expansion for
$\expv{\phi}$ but this is costly and unnecessary. The solution to
this loss of one order is to note that the first exponential
contains $S_0$ with a term linear in $\phi$ yet no factor of
$\delta$.  So if we were to differentiate with respect to the
variational source $j$ instead of the physical source $J$, we
bring down a sole factor of $\phi$. The problem is, of course,
that $S_0$ in $\Delta S$ carries $j$ dependence too, so direct
differentiation by the variational source would be of no use as it
would produce other terms. Yet, all is not lost, because we can
still use a trick to extract a factor of $\phi$ from $S_0$ without
interacting with the $\Delta S$ terms.

We notice that the $j$ coming from the $e^{- S_0}$ term in the
partition function is actually never seen explicitly in the full
development of the free energy density --- it is always hidden in
the $I_p$ integrals. On the other hand, the $j$ coming from the
$e^{\delta \Delta S}$ term appear only in factors of
\beq
 \Delta J:=J-j
 \label{DJdef}
\eeq
which are polynomial coefficients in the $\delta_1$ expansion of
the $J_p$ integrals.  It is easy then to keep $\delta J$ constant
while taking a partial differential with the remaining $j$ factors
by in effect only differentiating the $I_p$ factors with respect
to $j$
\begin{eqnarray}
\frac {\partial I_p(j,\Delta J)} {\partial j} & = & - I_{p + 1}
 \label{dpzdj}
\end{eqnarray}
To perform the $\frac {\partial f^{(R)}} {\partial j}$ operation,
we can go through the explicit expression of each $f_n$ (c.f.\
equation \eqref{eqf2numfinal}) and replace every occurrence of
$I_p$ by the right hand side of equation \eqref{dpzdj}. It is
straightforward to do the necessary replacements of the $I_p$ in
MAPLE and it produces optimised \verb|C| code for $\expv
{\phi}^{(R)} (\vecp,\vecv)$. Finally, for particular physical
values $\vecp$ we can substitute the appropriate variational
parameters $\barvecv$, which were determined by the relevant
optimisation condition (minimising free energy, PMS, FAC etc.).
Thus we can obtain $\bar {\expv {\phi}}^{(R)}
_\mathrm{expl}(\vecp) = \expv {\phi}^{(R)}_\mathrm{expl}
(\vecp,\barvecv)$.


Another physical quantity of interest is the susceptibility
$\chi$. It is defined by
\begin{equation}
\chi = - \pderivn {f} {J} {2}
\end{equation}
and is estimated in much the same way as $\expv \phi$ above. That
is to say, we will use both the direct numerical approach, and the
approach in which the individual orders of $\chi^{(R)}$ are
explicitly obtained. The direct numerical approach makes use of:
\begin{equation}
\chibar^{(R)}_\mathrm{diff} = \frac {\bar f^{(R)} (J + \epsilon) -
2 \bar f^{(R)}
  (J) + \bar f^{(R)} (J - \epsilon)} {\epsilon^2}
   \label{chidiffdef}
\end{equation}

Alternatively, finding the explicit cumulant expansion of $\chi$
proceeds by applying the recipe we introduced for calculating
$\expv \phi$:
\begin{equation}
\chi^{(R)} = \frac {\partial^2 f^{(R)}(j,\Delta J)} {\partial j^2}
\end{equation}
with $\Delta J$ defined in \tref{DJdef}. Writing $\chi^{(R)}$ as
an expansion, we identify the individual orders:
\begin{equation}
\chi^{(R)}_\mathrm{expl} (\vecp;\vecv)
 = \sum_{n = 0}^R \frac
{\delta^n} {n!} \chi_n \quad \Longrightarrow \quad \chi_n
 = \frac {\partial^2 f_n(j,\Delta J)} {\partial j^2}
\end{equation}
As for the calculation of $\expv \phi$, we employ Maple to form
optimised routines for $\chi^{(R)}_\mathrm{expl} (\vecp;\vecv)$.
The optimum value $\chibar^{(R)}$ is given by substituting the
appropriate set of physical parameters and the values for the
variational parameters obtained by optimising the free
energy\tnote{Evidence for lack of optimisation of other
quantities?}
\beq
\chibar^{(R)}_\mathrm{expl} (\vecp) :=
 \chi^{(R)}_\mathrm{expl} (\vecp;\bar{\vecv}) .
 \label{chiexpldef}
\eeq

To summarise, we have identified two different approaches for
evaluating some physical quantity $Q$ (once the free energy
density $f^{(R)}$ has been used to fix the variational
parameters).  We can estimate $Q$ directly using the optimised
free energy density $\bar{f}^{(R)} (\vecp)$, taking differentials
of this function. Variational parameters lose significance once
they have been used to optimise $f^{(R)}$ (c.f.\ equation
\eqref{phidiffdef}). Alternatively we produce an explicit
expansion for $Q^{(R)} (\vecp; \vecv)$. Some scheme is then needed
to produce the optimised values of the variational parameters
$\barvecv$, which could be derived either from the $Q^{(R)}
(\vecp; \vecv)$ expression itself in which case the optimised
values vary with the quantity $Q$ of interest.  Alternatively the
values which optimise the equivalent free energy expression might
be used.  However the optimised values $\bar{\vecv}$ are obtained,
they are used to give the estimated result for $Q$, $\bar{Q}^{(R)}
(\vecp) := Q^{(R)} (\vecp; \bar{\vecv})$. One of our aims will be
to produce results using both methods and to compare
them.\tnote{Note that in the PMS and FAC optimisation schemes, the
optimised parameters vary with the quantity $Q$ under discussion
as for each quantity optimise the explicit expression for the
quantity $Q^{(R)} (\vecp;\vecv)$, ignoring the free energy
completely. We will see that this fails to work in some simple
cases and will not pursue it at length.}

\section{Results}\label{sres}

\subsection{Ising Model}

We will start our discussion of results by looking at the Ising
model since this model reproduces most of the behaviour spin-1
model and of the full $\phi^4$ model with the benefit of being
numerically less complex. This numerical simplification comes from
two aspects: the restricted degrees of freedom, $\phi_i = \pm 1$,
and the reduced number of variational parameters, $\vecv = ( j )$.
The trial action is defined with only one variational parameter
because ultra-local quadratic and quartic terms are constant in
the Ising model.\tnote{Note, however, that a constant (Marko used
a quadratic) term in the trial action could conceivably contribute
to the expanded and truncated theory, a case which we will later
check.} The definition of $S$ for the Ising model \tref{Slatdef}
leaves the model depending on two physical parameters $\vecp =
(\kappa, J)$. We will refer to $\kappa$ as the \emph{inverse
temperature}\footnote{In the literature, the symbol $\beta$ is
predominantly used to denote the inverse temperature of an Ising
model. We use $\kappa$ to avoid a notation clash with the critical
exponent $\beta$.}.

The restriction of the field to $\pm 1$ reduces the $I_p$ factors
to simple functions\tnote{Are these meant to be normalised?}
\beq
 I_p = \left\{ \begin{array}{cl} 2 \cosh ( j ) & $p$ \mbox{ even}
 \\
 -2 \sinh ( j ) & $p$ \mbox{ odd}
 \end{array}
 \right.
 \label{isip}
\eeq

\subsubsection{Fixing the Variational Parameter} \label{secresOP}

Let us first consider the case with no physical source $J$. Then,
as in \cite{EIM}, we find that the linear variational parameter,
$j$, provides allows symmetry breaking to occur in this lattice
LDE approximation. As long as $J=j = 0$, the trial action is
invariant under the transformation of $\phi_i \rightarrow -
\phi_i$. Once $j \neq 0$, the symmetry no longer holds, unless it
is accompanied by the transformation $j \rightarrow - j$.  For
instance the free energy for fixed physical parameters
$\vecp=(\kappa,J=0)$ is an even function of $j$.  Thus $j=0$ is
guaranteed to be a turning point in variational space for $J=0$
problems.\tnote{Although we are discussing the Ising model at the
moment, this is quite true of the general formalism we developed,
i.e.\ it holds for the spin-1 and $\phi^4$ models.} This behaviour
is illustrated in figure \ref{fisf7j}, which shows plots of
$f^{(7)} (\kappa, J = 0; j)$ vs.\ $j$ for various values of
$\kappa$.\tnote{*** Get me order 7 examples of Ising model.}
\begin{figure}[!htb]
\begin{center}
\includegraphics{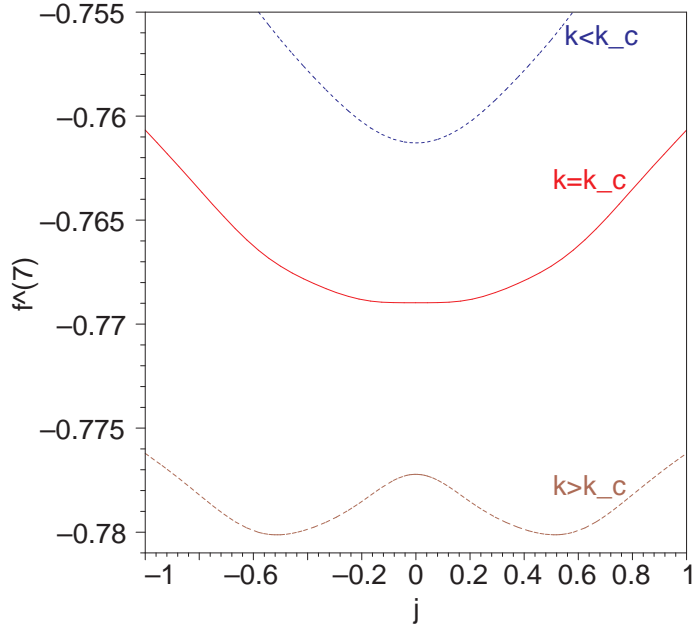}
\end{center}
\caption{Plot of $f^{(7)} (\kappa, J = 0; j)$ vs.\ $j$ for three
different values. These are $\kappa = \kappa_c - 10^{-2}$, $\kappa
= \kappa_c$, and $\kappa = \kappa_c + 10^{-2}$. The free energy is
of the 3D Ising model at 7th order.} \label{fisf7j}
\end{figure}
The turning point at $j=0$ need not be the unique minimum but what
we find is that the free energy profile has a simple form as a
function of $j$.  It is straightforward to apply our principle of
lowest free energy to choose the optimal variational parameter
value $\jbar$.  When $\kappa < \kappa_c$, the function $f^{(7)}$
indeed has a unique minimum at $j = 0$. For $\kappa \ra \kappa_c$,
we find the region around $j = 0$ flattening, and finally for
$\kappa > \kappa_c$ the free energy displays two clear minima,
either side of $j = 0$ with the $j=0$ becoming a maximum. Both
these minima are equally valid optimum points for $j$ as
guaranteed by the $Z_2$ symmetry of the model ($J=0$).
Numerically, the simplex optimisation method will choose one of
the two minima, depending on the initial step of the routine.

We can already make several deductions.  The optimal value $\jbar$
is a continuous but not smooth function of $\kappa$.  This
behaviour immediately tells us that our optimal value for the free
energy will not be a smooth function of the physical variable
$\kappa$.  Thus we clearly see that there will be some sort of
phase transition at $\kappa_c$ and the variational source is
itself a good order parameter even though it is not a physical
observable.

Also note that for the Ising model we find good global free energy
minima for both odd and even orders.  We will comment further on
this later.

\subsubsection{Non-zero Physical Source}\label{secresPSaFM}

For various calculations, we will require the behaviour of the
optimised free energy as a function of the physical source $J$ in
the region of  $J=0$.  Turning on the physical source, we break
the $Z_2$ symmetry and `skew' the free energy vs.\ $j$ curves.  In
the broken phase, one of the two minima acquires a lower free
energy than the other, depending on the sign of the source. This
effect can be seen in figure \ref{figresfvsjvarJ}, where we plot
the free energy curves with the system in the broken phase. The
inverse temperature is $\kappa = \kappa_c + 0.03$, and we use
three different values for the physical source.
\begin{figure}[!htb]
\begin{center}
 \includegraphics{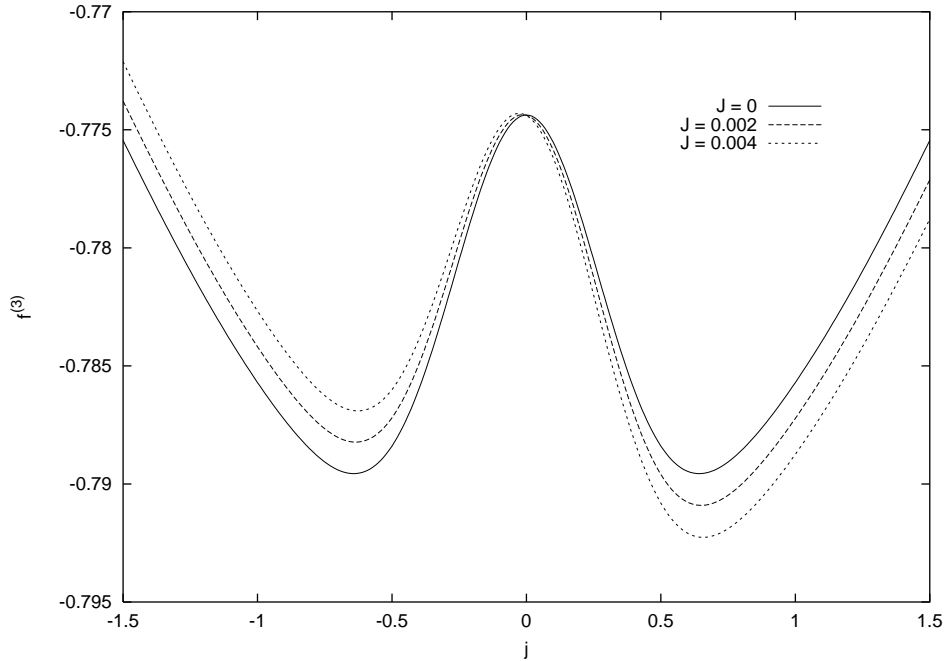}
\end{center}
 \caption{Plots of $f^{(3)} (J; j)$ vs.\ $j$ for a 3D Ising model,
at 3rd order. The inverse temperature is $\kappa = \kappa_c +
0.03$, i.e.\ the system is in the broken symmetry phase. The three
curves correspond to three different values of $J$. Each curve
consists of 300 calculated points, which are connected by straight
lines. The lines appear smooth due to the fine resolution of the
data points. }
 \label{figresfvsjvarJ}
\end{figure}
These are $J = 0$, $J = 0.002$ and $J = 0.004$.\tnote{Note that,
although not illustrated in the figure, it is obvious that by
switching the sign of the source, the free energy curve would skew
to the other side, the result of the $Z_2$ symmetry. The minimum
which previously had a higher free energy would become the more
favourable of the two.} With one local and one global minima, both
in deep wells, we had to be careful to select the global minimum
numerically.

\subsubsection{Critical Point} \label{secresIMCP}

The critical point, $\kappa_c$, was identified by searching for a
small interval where $\jbar(\kappa_-)=0$ and $\jbar(\kappa_+)\neq
0$ so that $\kappa_c= \half(\kappa_+ + \kappa_-) \pm
\half(\kappa_+ - \kappa_-)$.\tnote{Is this how we did it? Can you
give me the error?  If we have it we MUST quote it.}  In practice
because the free energy is insensitive to changes in the
variational parameter $j$ (it is a minimum in $j$ space) the
minimum in variational parameter space is located to a much lower
accuracy (in terms of the values for the variational parameters)
as compared to the numerical accuracy of the routines.  We need to
know the variational parameters in order to find the critical
point so uncertainties in these limit our knowledge of the
critical point.\tnote{MUST check this paragraph is correct.} This
is discussed in detail in appendix \ref{abhpa}.

In table \ref{tiskc}
\begin{table}[!htb]
\begin{center}
\begin{tabular}{c|c|c|c}
Order & $\kappa_c$ for 2D & $\kappa_c$ for 3D & $\kappa_c$ for 4D \\
\hline
1 & 0.2500000000000000 & 0.16666666666666667 & 0.1250000000000000 \\
2 & 0.3333333333333333 & 0.20000000000000000 & 0.1428571428571429 \\
3 & 0.3461538461538462 & 0.20270270270270270 & 0.1438356164383562 \\
4 & 0.3768115942028986 & 0.20963172804532578 & 0.1464393179538616 \\
5 & 0.3824833702882483 & 0.21006903118305165 & 0.1465228381635412 \\
6 & 0.3944031482291211 & 0.21343833354502731 & 0.1475861410791982 \\
7 & 0.3954564542557659 & 0.21353497619553782 & 0.1475854241672118
\end{tabular}
\end{center}
\caption{Numerical results for the critical points in the 2D, 3D
and 4D Ising models, at all orders up to 7. Accuracy is at least
to the last digit quoted (16th).} \label{tiskc}
\end{table}
we show the results obtained by searching for the critical inverse
temperature at all orders of the 2D, 3D and 4D Ising models.
Comparing these numerical values for the critical $\kappa$ to
those found exactly (see below in table \ref{kapcjexact}), we find
the numerics were accurate to a fractional error of $O(10^{-16})$,
i.e.\ to the accuracy quoted. This provides a further check of our
numerics.  A more detailed discussion of the errors in numerical
identification of the critical point is given in appendix
\ref{abhpa}.

In figure \ref{figresIM3DCrit} we plot the values of $\kappa_c$
for all the orders. We also indicate the result $\kappa_c =
0.221654$ obtained by Monte Carlo methods for the same model
\cite{FL,TB,HPV}. The figure shows that if \emph{only} the odd
orders are considered, an imaginary line going through the points
seems to converge to a value close to the one predicted by Monte
Carlo.
\begin{figure}[!htb]
\begin{center}
\includegraphics{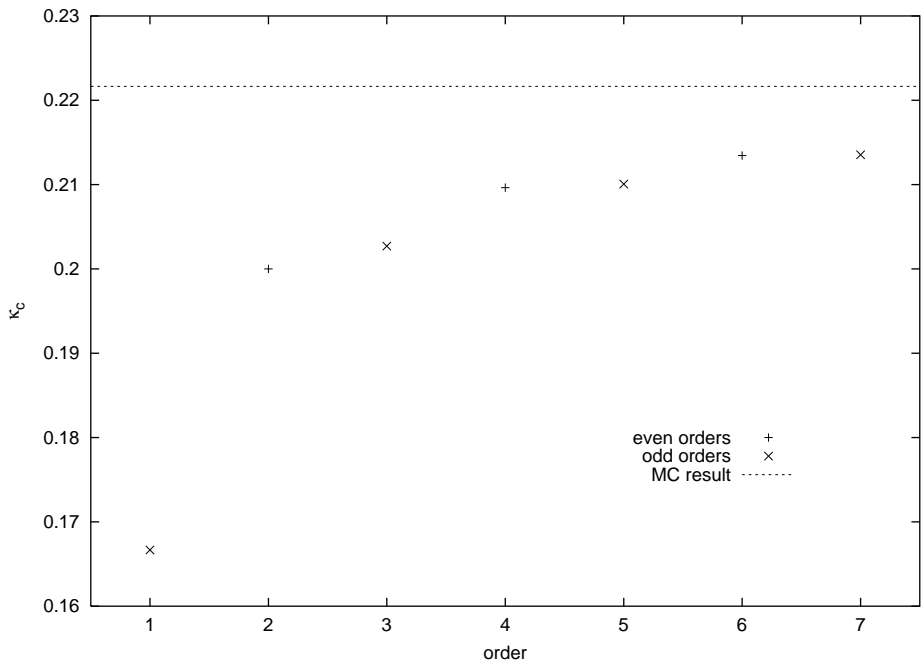}
\end{center}
\caption{The plot shows the values of $\kappa_c$ for the 3D Ising
  model, for all orders up to and including the 7th order. The Monte
  Carlo result is $\kappa_c = 0.221654$ \cite{FL,TB,HPV}.}
\label{figresIM3DCrit}
\end{figure}
The even points also show a smooth progression and appear to be
giving reasonable behaviour. However, in slightly more complicated
models the even orders do not give critical points at all.
Experience tells us to distrust the even orders of our model and
we will concentrate on the odd orders only.

\subsubsection{Field Expectation Value}
\label{secresFEV}

Consider first the evaluation of the expectation value via direct
numerical differentiation of the optimised free energy, $\expv
{\phibar}^{(R)}_\mathrm{diff}$ of \tref{phidiffdef}.  We present
the results for orders 3, 5 and 7 in figure \ref{fisphires}.
\begin{figure}[!htb]
\begin{center}
\includegraphics{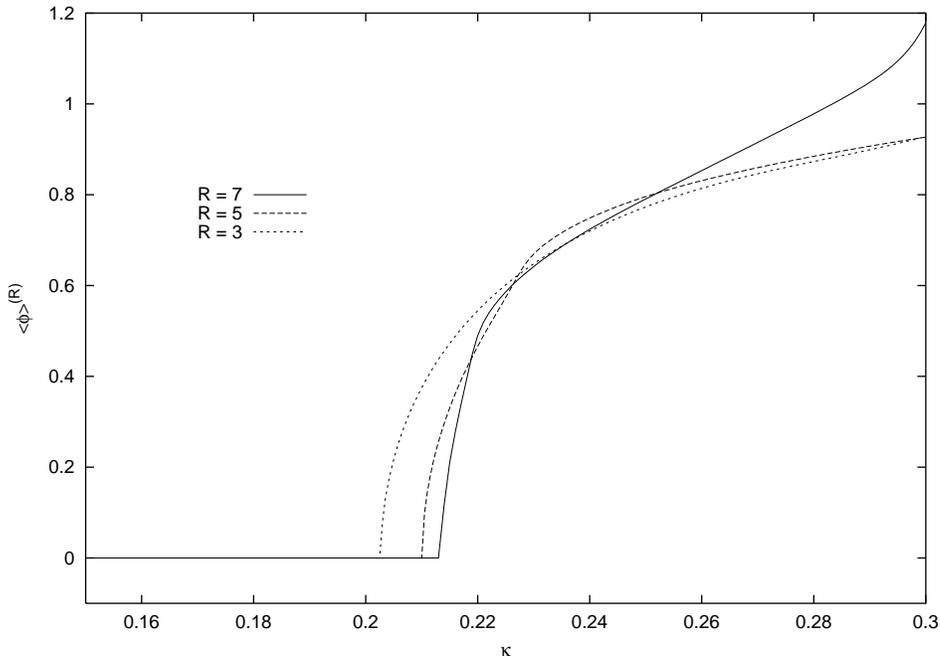}
\end{center}
\caption{The plot shows $\expv {\phibar}^{(R)}_\mathrm{diff}
(\kappa, J = 0)$ vs.\ $\kappa$ for $R = 3, 5, 7$, in the 3D Ising
model. The physical source $J = 0$. The expectation value is
calculated by direct numerical differentiation as given by
equation \eqref{phidiffdef}. Each curve consists of 400 calculated
points, which are connected by straight lines. The lines appear
smooth due to the fine resolution of the data points. }
\label{fisphires}
\end{figure}
We see that the expectation value displays similar behaviour for
the three odd orders considered and qualitatively, the results are
good. The expectation value clearly indicates the unbroken
symmetry phase with a vanishing value. For $\kappa > \kappa_c$, we
see the effect of spontaneous magnetisation. The only appreciable
difference between the three orders is the position of the
critical point.\tnote{Finally, for $\kappa$ going deeper into the
broken phase, the expectation value starts blowing up, which is
physically incorrect. The fact that the field expectation value
tends to blow up as $\kappa$ goes deeper into the broken phase
indicates a breakdown of the method.} Overall, the plot
demonstrates that the LDE can be used to model spontaneous
symmetry breaking with the expectation value as an order
parameter. We shall look later at the near-critical behaviour,
when we analyse the critical exponents.

In figure \ref{fisphirexpl} we have shown a plot of the
expectation value calculated by optimising the free energy with
respect to $j$ first for a given $\kappa$ and $J=0$, but then
substituting the optimal variational parameter $\bar{j}$ into the
explicit expansion of $\expv {\phi}^{(3)}_\mathrm{expl} (\kappa, J
= 0; j)$ to produce $\expv {\phibar}^{(3)}_\mathrm{expl} (\kappa,
J = 0)$  (see equation \tref{phiexpldef}). In figure
\ref{fisphirexpl} we show a plot of $\expv
{\phibar}^{(3)}_\mathrm{expl}$ and $\bar j$ for a range of
$\kappa$.
\begin{figure}[!htb]
\begin{center}
 \includegraphics{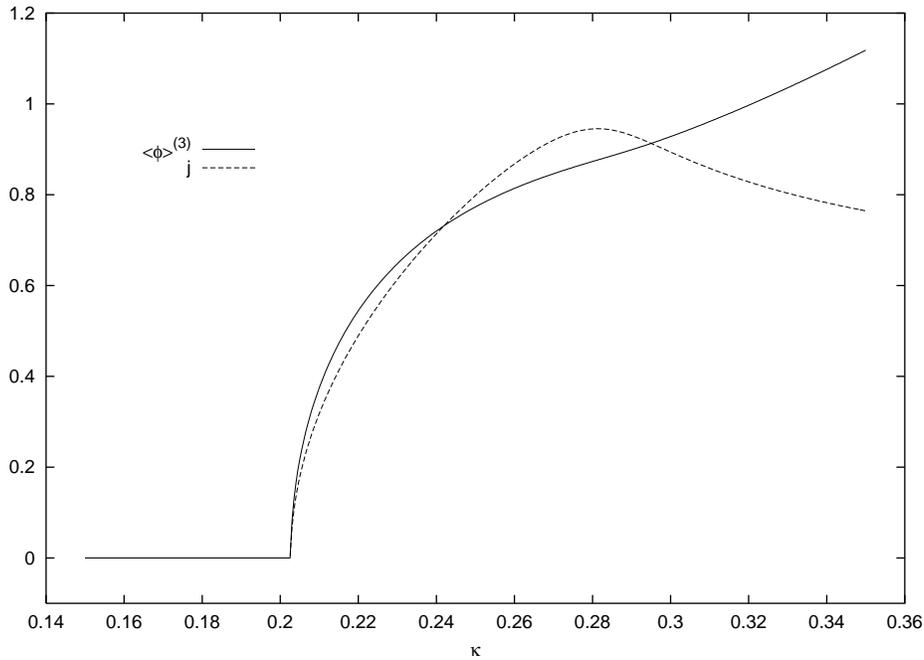}
\end{center}
\caption{Plot of $\expv {\phibar}^{(3)}_\mathrm{expl}$ and $\bar
j$ vs.\ $\kappa$ for the 3D Ising model at 3rd order, with the
physical source $J = 0$. Each curve consists of 300 calculated
points, which are connected by straight lines. The lines appear
smooth due to the fine resolution of the data points.}
 \label{fisphirexpl}
\end{figure}
These are results from a 3D, 3rd order run of the Ising
model,\tnote{Get me 7th order.} again with the physical source $J
= 0$. In broad terms, from the curve of $\expv
{\phibar}^{(3)}_\mathrm{expl}$ we clearly see again a phase
transition at a point which we denote by $\kappa_c$, the critical
inverse temperature.  The critical points are the same for both
approaches.

By clearly acquiring a zero and non-zero value, depending on the
phase, the expectation value is an order parameter of the model.
Additionally, we find that the same is true for the variational
parameter $j$. Thus we can use either order parameter to find the
value of $\kappa_c$ by searching for the inverse temperature at
which $\expv {\phibar}^{(R)}$ or $\bar j$ switch from a zero to a
non-zero value.

Note that the 3rd order curves using both methods look the same
for both methods.\tnote{Can I have optimisation attempts on phi
itself, i.e. a true Hugh style PMS?  Or at least an illustration
of why this fails.} In fact, accurate numerical comparison of the
two methods reveal differences of $\sim 10^{-20}$. Since we used
$\epsilon = 10^{-20}$ to calculate the numerical derivative (see
equation \eqref{phidiffdef}), this discrepancy is of purely
numerical nature.\footnote{The same numerical calculation was done
for the 5th and 7th orders too.} It is remarkable that the two
methods produce the same results, especially considering how
different the results will be for the susceptibility, discussed in
the next section.\tnote{Data please. Lets give the fractional
difference (at least in broken phase and for difference in
critical kappa.)}

\subsubsection{Susceptibility}
\label{secresS}

As with the field expectation value, the susceptibility is another
quantity that we can access by using optimised variational
parameters or by direct numerical differentiation of the free
energy density. The susceptibility is given by the second
derivative:
\begin{equation} \label{Susc}
\chibar^{(R)}_\mathrm{diff} = - \pderivn {\bar f^{(R)}} {J} {2}
\approx - \frac {\bar f^{(R)} (J + \epsilon) - 2 \bar f^{(R)} (J)
+ \bar f^{(R)} (J - \epsilon)} {\epsilon^2}
\end{equation}
where the right hand side defines the direct numerical
differentiation approach. The other approach is based on a
function $\chi^{(R)}_\mathrm{expl} (\kappa, J; j)$, which will
assume the `correct' value once an optimised $\bar j$ is
substituted, i.e.\ $\chibar^{(R)} (\kappa, J) = \chi^{(R)}
(\kappa, J; \bar j)$.

Figure \ref{figresIM3DChiCurves} shows a plot of four different
$\chibar^{(R)} (\kappa, J = 0)$ curves. Three of those are
$\chi^{(R)}_\mathrm{expl} (\kappa, J)$ calculated by optimising
$j$ first from the free energy, and then substituting this
optimised value into $\chi^{(R)}$, for $R = 3$, $R = 5$ and $R =
7$. The fourth curve shows $\chibar^{(3)}_\mathrm{diff} (\kappa, J
= 0)$ as calculated through direct differentiation \tref{Susc}.
This plot leads to important conclusions about the methods used to
calculate each curve, and we shall discuss these in what follows.
\begin{figure}[!htb]
\begin{center}
\includegraphics{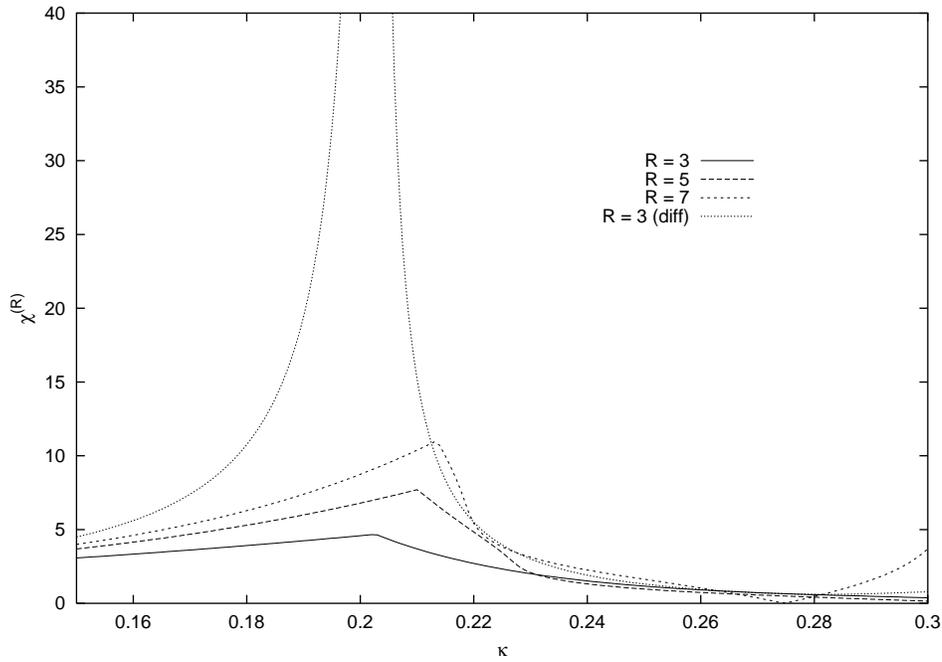}
\end{center}
\caption{The plot shows $\chibar^{(R)}_\mathrm{expl} (\kappa, J =
0)$ vs.\ $\kappa$ for $R = 3, 5$ and $7$ in a 3D Ising model.
These three curves are calculated using a previously optimised
$j$. The `$R = 3$ (diff)' curve is $\chibar^{(3)}_\mathrm{diff}
(\kappa, J = 0)$ calculated using direct differentiation, as given
by equation \eqref{Susc}. Each curve consists of 200 calculated
points, which are connected by straight lines. The lines appear
smooth due to the fine resolution of the data.}
\label{figresIM3DChiCurves}
\end{figure}

The curve denoting the direct numerical differentiation approach
$\chibar^{(3)}_\mathrm{diff}$ encapsulates the correct qualitative
behaviour of the susceptibility. It \emph{is} the case that $\chi$
blows up at the critical point, due to correlations extending over
the whole lattice (system).  This is due to the fact that this
approach is capturing the non-smooth nature of the field
expectation value apparent in figure \tref{fisphires}. The
$\chi^{(R)}_\mathrm{expl} (\kappa, J)$ form has only finite peaks
at all three orders, with the peaks getting larger as the order
increases. The fact that the peaks shift towards higher $\kappa$
is purely due to the value of $\kappa_c$ shifting at each
respective order (see table \ref{tiskc}). It is easy to see
(mathematically) why the peaks in the forms
$\chi^{(R)}_\mathrm{expl} (\kappa, J)$ become more pronounced at
higher orders. The explicit expressions for $\chi^{(R)}(\kappa,
J;j)$ are a sum of a finite number of terms, each a product of
$I_p$ factors, physical and variational parameters.  Like the
expressions obtained in a high temperature expansion on the
lattice, these expressions can not become infinite. Physically,
this is a reflection of the fact that at, for example, 3rd order,
our method can only correlate lattice sites spaced no more than
three sites apart. Diagrammatically, at 3rd order, two sites $i$
and $j$, sitting three sites apart from each other can be
correlated only through the diagram: \vspace{0.5cm}
\begin{equation}
\setlength{\unitlength}{1pt}
 \begin{picture}(60,10)
 \put(0,5){$i$}
 \put(0,2){\circle*{4}}
 \put(0,2){\line(1,0){20}}
 \put(20,2){\circle*{4}}
 \put(20,2){\circle*{4}}
 \put(20,2){\line(1,0){20}}
 \put(40,2){\circle*{4}}
 \put(40,2){\circle*{4}}
 \put(40,2){\line(1,0){20}}
 \put(60,2){\circle*{4}}
 \put(59,5){$j$}
 \end{picture}
\end{equation}
If $i$ and $j$ happen to be further apart, the method will never
be able to correlate them. The situation for $R = 5$ and $R = 7$
is analogous: the method can only correlate lattice sites which
are within $R$ sites of each other and they can never see the
infinite correlation length near the critical point. This is the
reason why the $\chi^{(R)}_\mathrm{expl}$ peaks grow as orders
increase, but still remain finite for finite $R$.

\subsubsection{Critical Exponents}
\label{secresCE}


The $\beta$ critical exponent is defined in terms of the
\emph{reduced inverse temperature} $\kappa_r$\tnote{Try with
$\kappa_c$ on the denominator.}
\begin{equation} \label{betadef}
\lim_{\kappa \ra \kappa_c} \left(\expv{\phibar} \right)
 \propto
 \left| \frac {\kappa - \kappa_c} {\kappa} \right|^\beta
  = \left| \kappa_r \right|^\beta,
 \;\;\;
  \kappa_r  := \frac {\kappa - \kappa_c} {\kappa}
\end{equation}

Let us take the three-dimensional order seven results as a
prototypical example.  Using data produced with 155 digit
accuracy, we fit\tnote{See \cite{ZJ} (27.22).  The one in Marko's
thesis has the correct $\kappa_r$ with $\kappa$ or $\kappa_c$ on
the bottom.}
\begin{equation}
 \label{betafit}
 \expv{\phi}^{\mathrm(fit)} =  A \left|\kappa_r \right|^\beta \left| \ln (\kappa_r)
 \right|^c
\end{equation}
using 30 digit accuracy on the fitting for 100 values of $\kappa
\in \langle \kappa_c, \kappa_c + 10^{-10}]$.  The best fit is for
$\beta=0.5 + 3 \times 10^{-10}$ and
$(\kappa_c/\kappa_{c,\mathrm{exact}} -1)= 1.4 \times 10^{-20}$
with $\chi^2=8 \times 10^{-20}$ as figure \ref{fchi2bfit} shows.
The reference value $\kappa_{c,\mathrm{exact}}$ is given later in
table \ref{kapcjexact}.
\begin{figure}[!htb]
\begin{center}
\includegraphics{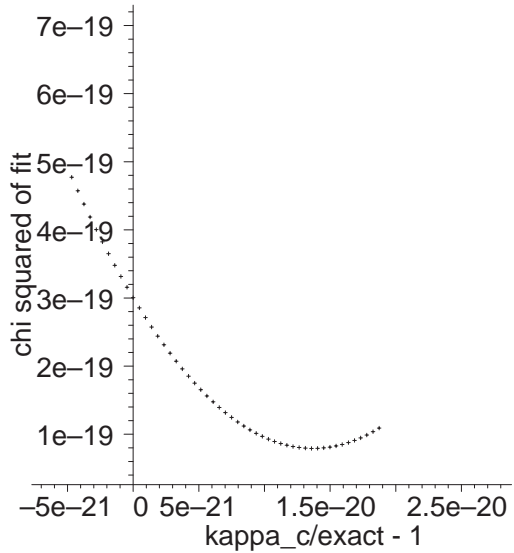}
\end{center}
\caption{Plot of the standard statistical measure $\chi^2$ for the
best fit of \tref{betafit} to the behaviour of the 3D Ising model
7th order results near critical point for a fixed
$(\kappa_c/\kappa_{c,\mathrm{exact}} -1)$.}
 \label{fchi2bfit}
\end{figure}
The fit is very good so the fact that $\beta$ is not exactly a
half looks significant at first given the high calculational
accuracy.

However the fractional differences between this best fit curve and
the data points are shown in figure \ref{fbetafit}.
\begin{figure}[!htb]
\begin{center}
\includegraphics{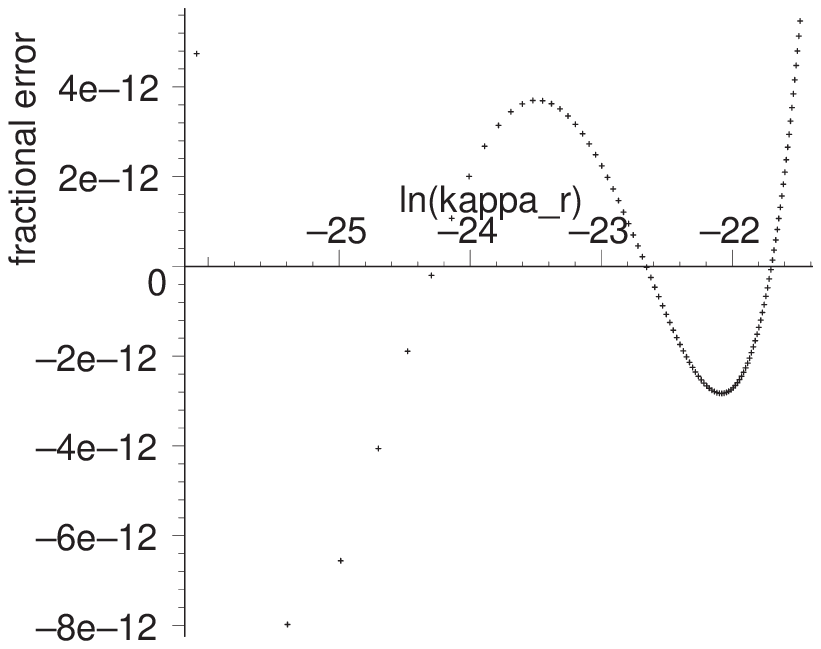}
\end{center}
\caption{For the best fit of the critical behaviour of the 3D
Ising model 7th order results ($J=0$), a plot of
$\ln({\expv{\phibar}}^{\mathrm{(fit)}} ) / (\expv {\phibar}^{(7)}
)$ vs.\ $\ln \left| \kappa_r \right|$ for $\kappa \in ( \kappa_c,
\kappa_c + 10^{-10} ]$ with $\beta=0.5 + 3 \times 10^{-10}$ and
$(\kappa_c/\kappa_{c,\mathrm{exact}} -1)= 1.4 \times 10^{-20}$.}
\label{fbetafit}
\end{figure}
This shows a small but significant systematic error that we have
not been able to eliminate, perhaps due to a poor choice for the
form of the fitting function \tref{betafit}.  There is no reason
why our approximate form should have exactly the same functional
form as the full solution and we are working here to high levels
of numerical accuracy. Fitting data in the same way but in the
ranges $\kappa \in \langle \kappa_c, \kappa_c + 10^{-r}]$ for
$r=6$ and $8$ reveals that the all the errors and the deviation of
the best estimate for $(\beta-\half)$ reduce by two orders of
magnitude as we use data ranges with $r=6,8$ and then $10$, closer
and closer to the critical point. If we fit a pure power law
($c=0$ in \tref{betafit}) then the fit is slightly worse, the beta
is different and again differs by the same order of magnitude from
$1/2$.  Attempts to fit forms for other corrections to
\tref{betadef} to the data produce similar results but no
consistency in any shift of $\beta$ from one half. Thus, despite
our high numerical accuracy, we conclude that the deviation from
$\beta=\half$ is not significant and that our data is consistent
with $\beta= 1/2$, again to about nine significant digits.


Power law behaviour of the susceptibility near the critical point
defines the $\gamma$ critical exponent as
\begin{equation} \label{gammaDef}
\chi \propto \left| \kappa_r \right|^{- \gamma}
\end{equation}
To find the value of $\gamma$, we used the susceptibility
$\chi^{(R)}_\mathrm{diff}$ calculated by direct differentiation of
the free energy density \tref{Susc}. Following our experience with
$\beta$ we merely fitted the simple power law form \tref{gammaDef}
to the data for $\ln \left| \chibar^{(3)} \right|$ with $\kappa$
taken in the broken phase and  $\kappa \in ( \kappa_c, \kappa_c +
10^{-8} ]$. The result is $\gamma = 1*(1\pm 10^{-8})$ for orders
3, 5 and 7.


Finally the $\delta$ critical exponent is defined by the following
power law
\begin{equation} \label{deltaDef}
\expv \phi \sim \left| J \right|^\frac {1} {\delta} \qquad
\rm{for} \ \kappa_r = 0
\end{equation}
Note that the $\delta$ critical exponent has nothing to do with
the bookkeeping parameter $\delta$ defined in the LDE methodology.
Again we fitted the power law form \tref{deltaDef} to the data for
${\expv{\phi}}_\mathrm{diff}$  with $\kappa=\kappa_c$. The result
of the fit  at orders 3, 5 and 7 is $\delta = 3*(1\pm 10^{-8})$,
to similarly high accuracies as for the other critical exponents.


A summary of the critical exponents for the Ising model, as found
by our LDE method is shown in table \ref{tabrescritexpsum}.
\begin{table}[!htb]
\begin{center}
\begin{tabular}{c|c|c}
Exponent & LDE & MC+HT \\
\hline
 $\beta$ & 0.5 & 0.3485 \\
 $\gamma$ & 1 & 1.3177 \\
 $\delta$ & 3 & 4.780 \\
 $\delta/(\gamma/\beta+1)$ & 1 & 1.000
 \end{tabular}
\end{center}
\caption{The values of the three critical exponents $\beta$,
$\gamma$ and $\delta$ for the Ising model in three dimensions.
First as obtained by our LDE study (at orders 3, 5 or 7) secondly
by a combination of Monte Carlo and high-temperature expansions
\cite{HR99,CHPPV}.} \label{tabrescritexpsum}
\end{table}
For comparison, we give the values of the same critical exponents
as found by \cite{CHPPV}, using a combination of Monte Carlo
simulations based on finite-size scaling methods, and
high-temperature expansions. We note that the values obtained by
the LDE match those predicted by mean field theory up to errors.

One final test is to check the universal relationship $\delta =
1+(\gamma/\beta)$ but as mean field values satisfy this, our
values also satisfy this requirement within the accuracy of our
measurements.

\subsubsection{The Ising model in other dimensions}
\label{secresOR}

So far we have given a detailed description of our results for the
3D Ising model. In table \ref{kapcjexact} we list critical inverse
temperatures for all orders of the 2D, 3D and 4D Ising models.
Again these are exactly solvable so can be used to check our
numerical calculations. The critical exponents can then be
measured as already described.  Again they are identical to the 3D
results, that is mean field values at orders 3, 5, 7 within the
accuracy of our calculations.  This independence of critical
exponents with dimension is also characteristic of mean field
theory but is not seen in the true results.

\subsubsection{The Ising model and other optimisation schemes}
\label{sisopt}

Once the complete expressions have been obtained for the various
scalar field models, it is relatively quick to compute them for a
range of values in the Ising model with its single variational
parameter and with $I_p$ `integrals' being given by simple
functions.  It is an ideal place to illustrate other aspects of
the LDE method.  We will just study the $J=0$ case.

First let us look at optimisation scheme known as FAC - fastest
apparent convergence.  In this scheme one chooses the variational
parameters such that the last term in the delta expansion (or
sometimes the last $r$ terms) is zero.  The idea is that in well
behaved series the last term can often be used to get an idea of
the error due to truncation of the series so the `optimal' result
is where this is zero.

The modulus of the fractional contribution to the free energy from
each order in the full delta expansion is plotted in figure
\tref{fisfc108d3o7} for the seventh order calculation.
\begin{figure}[thb]
\begin{center}
{\includegraphics{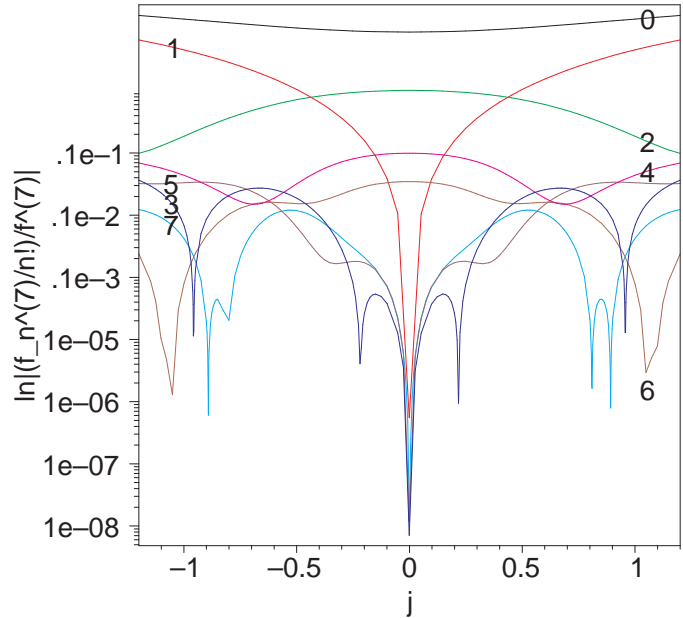}}
\end{center}
\caption{The eight terms in the delta expansion of the free energy
of the Ising model in three dimensions up to order seven with
$J=0$ and $\kappa = 0.23061777$. Normalised with respect to the
total free energy, the log of the absolute value is plotted. The
numbers on one end of every curve indicate the order of the term
plotted. Note that most spikes represent points where a term is
going to zero but numerical and plotting limitations mean the
curves are of finite extent. Likewise, the curves ought to be
exactly symmetric about $j=0$ but near the spikes numerical
limitations prevent this appearing on the plot.}
 \label{fisfc108d3o7}
\end{figure}
The spikes ought to be going to minus infinity on this log plot as
these are points where a term goes through zero.

It is clear that the odd terms always go to zero at $j=0$. This
should be expected as all the odd integrals $I_p$ are zero when
$j=0$, and at odd orders, for a trial action $S_0$ made up of only
odd powers of fields, every contribution in the sum of such
contributions making up the odd order expression must contain at
least one $I_p$ with $p$ odd.  Thus only an FAC procedure based on
odd terms alone, will provide a suitable unbroken $j=0$ solution.
Thus let us further confine ourselves to an FAC procedure where we
demand that the optimal variational parameters $\barvecv$ are such
that the last term in the expansion is zero
\beq
 f^{(R)}_\mathrm{FAC}(\vecp) := \sum_{n=0}^R \frac{\delta^n}{n!} f_n(\vecp,\barvecv), \;\;\;
 f_R (\vecp,\barvecv) = 0
 \label{FACRdef}
\eeq

Now we notice that, at least for the value of $\kappa$ used in
figure \tref{fisfc108d3o7}, there are in fact non-zero values of
$j$ which satisfy the FAC optimisation criterion \tref{FACRdef}.
There are four: two positive $j$ and the same solutions with
opposite sign (as there must be under the $Z_2$ symmetry at
$J=0$).  Further investigation shows that these two new solutions
appear first at $\kappa_{\mathrm{FAC},C}^{(7)}=\kappa_C^{(7)} *
(1.048 \pm 0.002)$ where $\kappa_C^{(7)} \approx 0.2135$ is the
critical $\kappa$ value found in the same model but using the
minimum free energy criterion \tref{iPMSdef}.  Now the problem for
$\kappa>\kappa_{\mathrm{FAC},C}^{(7)}$ is which minimum to choose
--- we have three distinct possibilities after symmetry is taken
into account.  This illustrates a general problem when using
criteria such as FAC and PMS.  They often present multiple
possible solutions for the optimal variational parameters and one
must use further criteria, sometimes no more than physical
intuition, to choose one solution.  The minimum free energy method
we use guarantees a unique answer, up to symmetry, except at
transition points.

To finish with FAC, let us choose the $j$ solution that has the
lowest free energy as well as satisfying \tref{FACRdef} for this
seventh order $J=0$ example.  As figure \tref{fisfc108d3o7} shows
this is the smallest non-zero $j$ solution (the free energies are
negative, so this has the largest free energy ratio when compared
to the $j=0$ solution).
\begin{figure}[thb]
\begin{center}
 {\includegraphics{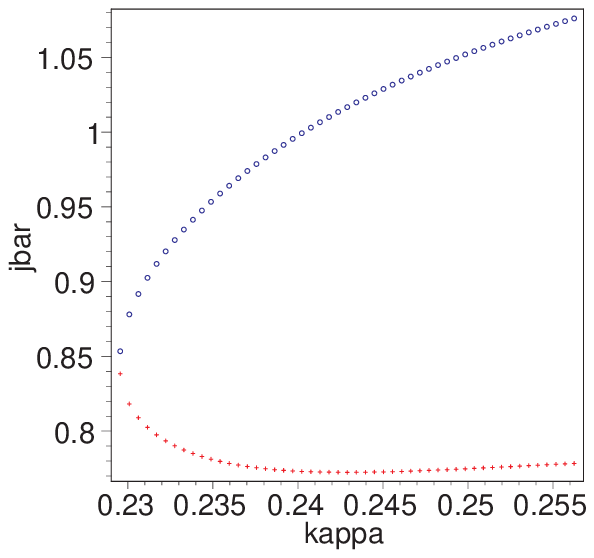}}
 \hspace*{0.5cm}
 {\includegraphics{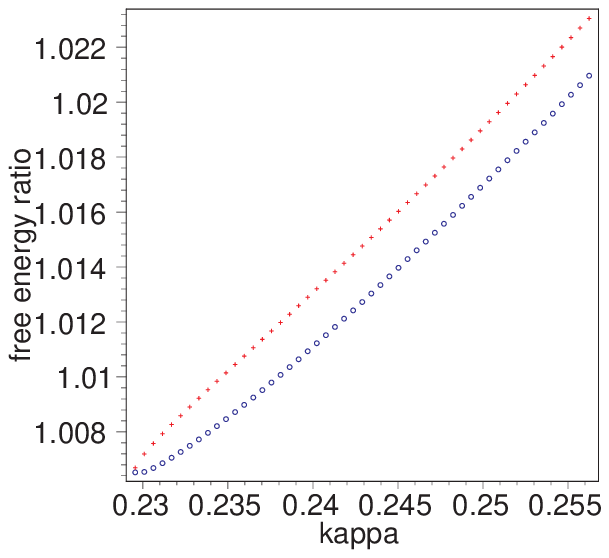}}
\end{center}
\caption{The optimal non-zero $j$ solutions (left hand plot) and
the resulting free energy normalised to the free energy at $j=0$
(right hand plot) for the Ising model in three dimensions at order
seven with $J=0$. Note the smaller $j$ solution (crosses) has the
most negative free energy.}
 \label{fisfacjkd3o7}
\end{figure}
This solution appears at about $\jbar\approx 0.85$ at $\kappa=
\kappa_{\mathrm{FAC},C}^{(7)}$ and so this FAC criterion is giving
us a \emph{first order transition}.  Any physical quantity we
calculate will suddenly change in value as we increase $\kappa$
through $\kappa_{\mathrm{FAC},C}^{(7)}$ as the $j$ parameter is
suddenly changed from $0$ to around $0.85$.  In some models this
is a good thing.  For instance in the Electroweak model for some
parameter ranges we expect a first order transition and this
sudden change in the optimal variational parameter solution can
give this behaviour in LDE as obtained in \cite{EJR98a,EJR98b}.
However, we are expecting a second order transition in this model.
Our conclusion is that FAC is an unsatisfactory optimisation
scheme for the optimised hopping parameter expansion of the Ising
model.

We next turn to PMS.  In this scheme, one looks for turning points
\emph{in the quantity of interest}.  Thus for a fixed set of
physical parameters $\vecp$  the optimal values chosen for the
variational parameters $\vecv$ will vary as we study different
quantities.  This is a standard procedure when PMS is applied to
issues such as scheme dependence in particle physics
\cite{PDG,St81}.

First let us apply PMS to the free energy. The plot in figure
\ref{fisf7j} shows how this works.  For $\kappa<\kappa_c$ there is
a single turning point at $j=0$ at the global minimum of the free
energy so the PMS gives the same answer as free energy
minimisation in the unbroken phase.  However for $\kappa>\kappa_c$
there are three turning points and in principle PMS does not
distinguish between turning points in the variational parameter
space.  In practice one often chooses the flattest turning point
(for least sensitivity to unphysical parameters) but here there is
little to choose between them. Failing this, with PMS one often
falls back on choosing the turning point which makes the most
physical sense. Here that might be the ones with the lowest free
energy or with $j \neq 0$ since we are expecting symmetry
breaking.  Of course with such additional arguments one selects
the same solution as we had when simply searching for the minimum
of the free energy.  In this sense PMS is working here as well as
the minimum free energy criterion. What it does highlight is that
PMS need not give a unique solution nor a single solution.

More interestingly we can try the PMS on other quantities. Let us
look at the expectation value of $\phi$ as a function of
variational $\vecv=\{ j \}$ and physical parameters $\vecp = \{
\kappa, J=0\}$.  This is shown in figure \ref{fisjphi}.
\begin{figure}[thb]
\begin{center}
 {\includegraphics{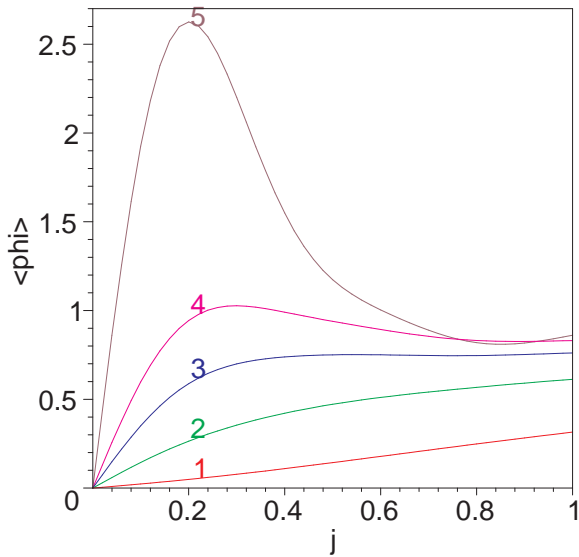}}
\end{center}
\caption{$\expv{\phi}(\kappa,j)$ in the three dimensional Ising
model at order seven with $J=0$ and fixed $\kappa$.  The labels
indicate that $(\kappa - \kappa_c) = -0.05$ (1), $0.0$ (2),
$+0.03$ (3), $+0.05$ (4) and $+0.1$ (5). Note that the new turning
points appear at a kappa about 14\% higher that $\kappa_c$, the
value shown with curve three.  The curves are all odd in $j$ so
negative $j$ are not displayed.}
 \label{fisjphi}
\end{figure}
The main point to note is that $\expv{\phi}$ is an odd function of
$j$ and for large ranges of $\kappa$, the low kappa regions, there
are no turning points in $\phi$.  For larger $\phi$'s there are
four turning points, with one nice shallow local minimum as a
function of $j$ (maximum for negative j) appearing at some $\kappa
\neq \kappa_c$.  However it is clear that the PMS  fails to give
reasonable qualitative behaviour for this quantity.

Thus for the Ising model, the FAC fails to work at all, while the
PMS fails when applied to some quantities, and works on
others\footnote{Presumably PMS works for those quantities even in
$j$.} if implemented with some extra reasonable but ad-hoc
criteria. On the other hand the unambiguous minimum free energy
principle gives great qualitative behaviour for all quantities
(when calculated using derivative of the optimised free energy)
but it produces poor critical behaviour of a type seen in the mean
field approximation.

\subsection{Spin-1 Model} \label{secresS1M}

The spin-1 model differs from the Ising model in that $\phi$ can
acquire the values $\pm 1$ and $0$. This means that a quadratic
ultra-local term is no longer redundant (as it was for the Ising
model), so $S$ and $S_0$ are (c.f.\ equation \tref{Slatdef}):
\bea
 S & =& - \kappa \sumnn {i}{j} \phi_i \phi_j
  + \sumn {i} \left[ J \phi_i + \alpha \phi_i^2
\right]
 \\
S_0 & =& \sumn {i} \left[ j \phi_i + k \phi_i^2 \right]
\eea
Thus the $I_p$ integrals \tref{Ipdef} are again simple functions
\beq
 I_p(j,k) :=
 \left\{ \begin{array}{rl}
 1 + 2 e^{-k}  \cosh (j) & (p =0)
 \\
 2 e^{-k}  \cosh (j) & (p \mbox{ even and positive})
 \\
 -2 e^{-k} \sinh (j) & (p \mbox{ odd})
 \end{array}
 \right.
 \label{Ipspin1}
\eeq
However, now we are optimising in the two-dimensional variational
space of $\vecv=(j,k)$. We set the parameter $\alpha = \ln 2$
\cite{HPV, BLH}.\tnote{Why???}

The optimisation with respect to two parameters causes no
additional problems at odd orders.  The $j$ parameter acts as an
order parameter as before.  At low values of $\kappa$ $k=1$ and
$j=0$ gives the lowest free energy, i.e.\ where the variational
parameters equal the physical parameters $\vecv = \vecp$ and the
trial action exactly equals the ultra-local part of the physical
action. However, the gradient in the free energy with respect to
$j$ slowly decreases and becomes zero at some point and this is
the critical point. For higher $\kappa$ values there are a pair of
minima at $j \neq 0$ and $k \neq 1$.\tnote{Can we plot the k
behaviour?} The free energy as a function of the variational
parameters is shown for the third order in figure \tref{fs1fd3o3}
for values of kappa either side of this critical point.
\begin{figure}[!htb]
\begin{center}
\includegraphics{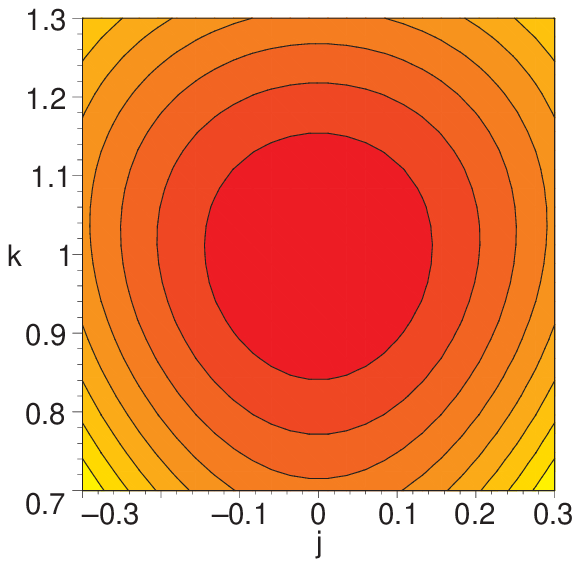}
 \hspace*{1cm}
\includegraphics{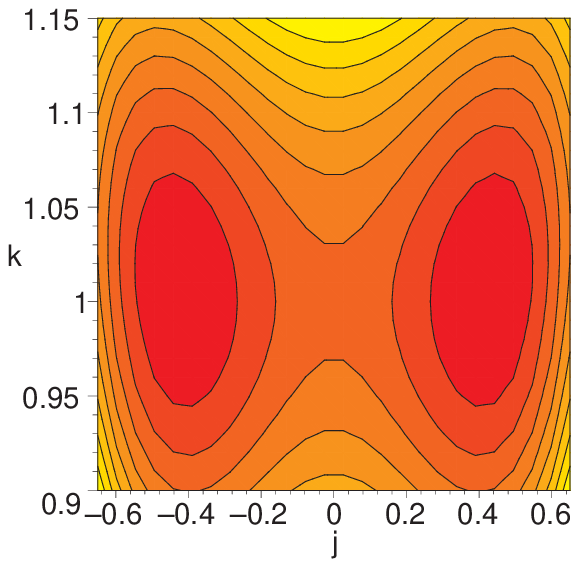}
\end{center}
\caption{Contours of constant free energy (highest values at edges
of plot) for the spin one model in three dimensions at order 3 for
$\kappa =0.9 \kappa_c$ (left) and $\kappa =1.1 \kappa_c$ (right).
Here $\kappa_c$ is a genuine critical point.} \label{fs1fd3o3}
\end{figure}

Table \ref{tabCrit2-4D_S1_jk} summarises the critical inverse
temperatures found for the 2D, 3D and 4D cases.  Exact expressions
can be obtained but they involve integer powers of $1/e$ and are
not given here.\footnote{This is because in the unbroken phase,
the coefficient of the quadratic term $k$ prefers to be equal to
that in the physical action, here chosen to be 1.  Thus the $I_p$
integrals contain $1/e$ factors.}\tnote{Only checked this for d=3
order 3.}
\begin{table}[!htb]
\begin{center}
\begin{tabular}{|c|c|c|c|}
\hline
Order & $\kappa_c$ for 2D & $\kappa_c$ for 3D & $\kappa_c$ for 4D \\
\hline \hline
1 & 0.5000000000000000 & 0.33333333333333333 & 0.2500000000000000 \\
3 & 0.5753424657534247 & 0.36464088397790055 & 0.2670623145400593 \\
5 & 0.6060227414719480 & 0.37172944976396613 & 0.2697296215999957 \\
7 & 0.6475789767717944 & 0.37519182797115188 & 0.2700940775993100 \\
\hline
\end{tabular}
\end{center}
\caption{List of all critical points for the 2D, 3D and 4D spin-1
models, at all odd orders up to 7. The values are obtained by
using the linear and quadratic variational parameters, $j$ and
$k$. The result obtained by Monte Carlo methods for the 3D case is
$\kappa_c = 0.383245$ \cite{HPV}. } \label{tabCrit2-4D_S1_jk}
\end{table}

The critical exponents behave in the same way as for the Ising
model: $\beta = 0.5$, $\gamma = 1$ and $\delta = 3$ to over eight
significant digits, irrespective of order or dimension.

We present only odd orders, as the even orders do not produce good
behaviour in the variational space.\tnote{Confirmed with Maple.
Exemplary 6th and 7th order plots please. Contour plot. Tim can do
on MAPLE.} It is the behaviour in the quadratic variational
coefficient, $k$ which leads to the even orders failing to produce
good minimum energy values. For instance in figure
\tref{fs1fkjsecd3o2} the free energy of the spin one model at
second order is shown.  The free energy behaves reasonably as a
function of $\kappa$ and we can identify a value $\kappa_c$ where
where the second derivative with respect to $j$ is zero at
$j=0,k=1$.  However, there, as for all values shown the free
energy has no turning point with respect to $k$ and no obvious
minimum.
\begin{figure}[thb]
\begin{center}
{\includegraphics{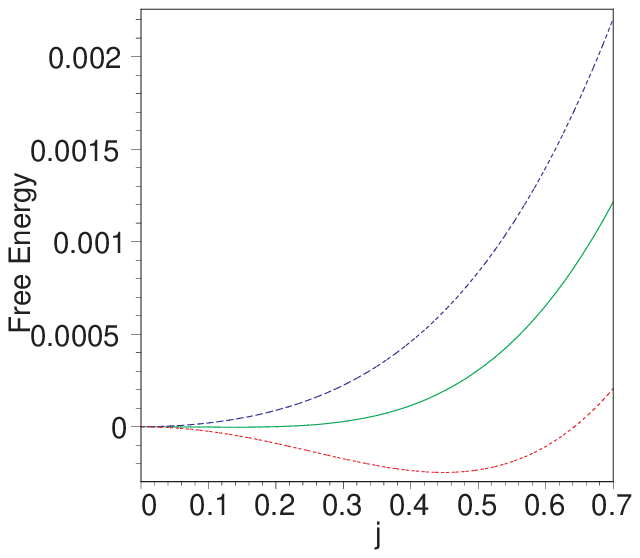}}
 \hspace*{1cm}
{\includegraphics{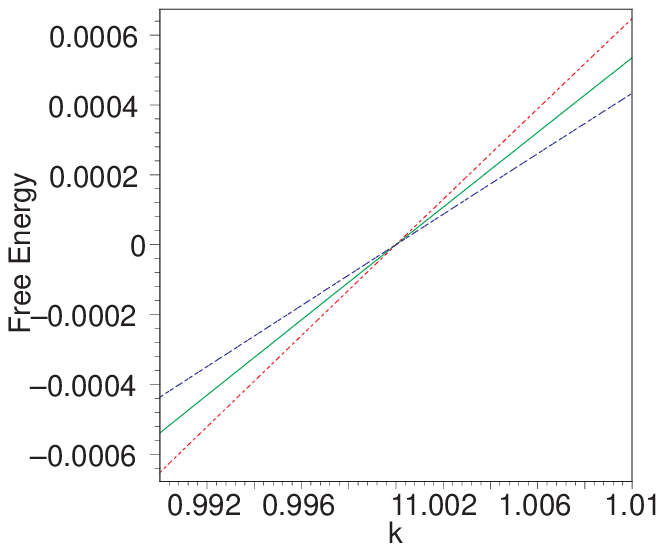}}
\end{center}
\caption{The free energy of the spin one model in three dimensions
at order 2 at $\kappa/\kappa_c=$ 0.9(top blue curve), 1.0 (middle
green continuous curve), 1.1 (bottom red dashed curve). At order
2, $\kappa_c$ is the point where the second derivative of the free
energy with respect to $j$ is equal to zero at $j=0,k=1$, as the
left hand graph indicates (its for $k=1$). The figure on the right
is at $j=0$ and shows that it is the lack of a turning point in
the $k$ variational parameter which prevents the method working at
order 2. (Tim Maple analysis)}
 \label{fs1fkjsecd3o2}
\end{figure}

\begin{figure}[thb]
\begin{center}
{\includegraphics{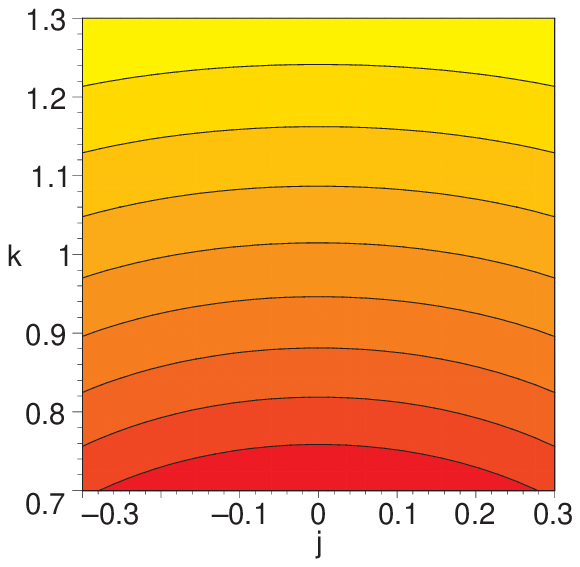}}
 \hspace*{1cm}
{\includegraphics{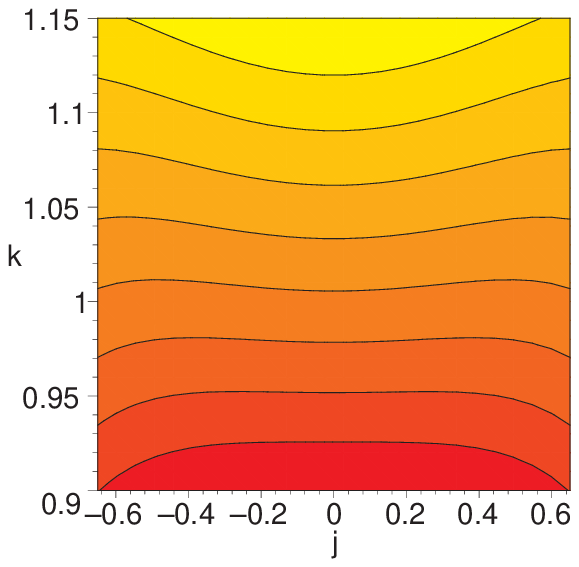}}
\end{center}
\caption{The free energy of the spin one model in three dimensions
at order 2 at $\kappa =0.9 \kappa_c$ and $\kappa =1.1 \kappa_c$.
Here $\kappa_c$ is the point where the second derivative of the
free energy with respect to $j$ is zero at $j=0,k=1$.  It shows
that it is the behaviour with respect to the $k$ variational
parameter which stops the method working at order 2. (Tim Maple
analysis)}
 \label{fs1fkjcontd3o2}
\end{figure}

Interestingly, we tried reducing the number of variational
parameters to one, namely, we used $j$ only. The results produced
in this way were the same as with the $k$ variational parameter
included, for all cases. The only differences occurred at even
orders, at which we could not find any reliable results anyway.

\subsection{$\phi^4$ Model} \label{secresPhi4M}

As a reminder, the action and the trial action of the $\phi^4$
model are given by
\begin{eqnarray}
S & =& - \sumnn{i}{j} \phi_i \phi_j + \sumn{i} \left[ J \phi_i +
\alpha \phi_i^2 + g \phi^4_i \right]
\\
L_0 (\phi_i,\vecv) & :=& \left[ j \phi_i + k \phi_i^2 + l \phi_i^4
\right]
\end{eqnarray}
Thus the $I_p$ integrals \tref{Ipdef} are now non-trivial and we
are optimising in the two-dimensional variational space of
$\vecv=(j,k)$ if we set $l=g$ or more generally we work with a
three dimensional variational space $\vecv=(j,k,l)$. Note that
here the inverse temperature $\kappa$ has been scaled to unity
when discretizing the $\phi^4$ action (c.f.\ section \ref{slsfm}).
Instead, we use $\alpha$, the lattice \emph{mass parameter}, as
the physical basis for `driving' the system through the phase
transition, so we have physical parameters $\vecp=(J,\alpha,g)$.
We set $g = 25$, to agree with \cite{EIM, WZZSDYX}.\tnote{I want
an order 7, $g=1/25$ run, perhaps with a
$\phibar_\mathrm{diff}^{(7)}$ and exemplary $\beta$ value.}

The crucial difference in this model as compared to the Ising and
spin-1 models, is the number of degrees of freedom of the field.
For the $\phi^4$ model, the field is a continuous parameter, and
the trace in the partition function becomes an integral.
Therefore, the numerical calculation of the $I_p$ factors is a
computationally intensive task as they are now integrals whereas
they were standard functions for the previous two models. We had
to make sure that the accuracy of our numerical integration was
appropriate given the accuracy problems encountered in
manipulating the long expressions the LDE gives for the quantities
of interest. For instance, when 256-bit arithmetic was used (which
gives approximately 77 decimal places of accuracy), we ensured
that the integrals were always evaluated to a precision of at
least 70 decimal places.

Table \ref{tabCrit2-4D_Phi4_jk} summarises the critical values of
$\alpha$ at odd orders for the 2D, 3D and 4D cases.
\begin{table}[!htb]
\begin{center}
\begin{tabular}{|c|c|c|c|}
\hline
Order & $\alpha_c$ for 2D & $\alpha_c$ for 3D & $\alpha_c$ for 4D \\
\hline \hline
1 & -14.585428484653014 & -10.007548118296870 & -6.956574606584720 \\
3 & -18.407621223877752 & -11.560051936952422 & -7.887721708383015 \\
5 & -20.106877371919612 & -11.918151992492446 & -8.018314178185453 \\
7 & -21.240866931195218 & -12.138250831659820 & -8.062177866626536 \\
\hline
\end{tabular}
\end{center}
\caption{List of critical $\alpha_c$ for the 2D, 3D and 4D
$\phi^4$ model, at all odd orders up to 7. The values are obtained
by using the linear and quadratic variational parameters, $j$ and
$k$.} \label{tabCrit2-4D_Phi4_jk}
\end{table}
The critical values were calculated using only the linear and
quadratic variational parameters. The critical exponents $\beta$,
$\gamma$ and $\delta$ turned out to be 0.5, 1 and 3, respectively,
again reproducing the values found in the Ising and spin-1 models
to at least eight significant figures.

The 3rd order result in 4D successfully reproduces and refines the
result $\alpha_c = -7.88$ found by a similar variational cumulant
approach in \cite{WZZSDYX}. Two other studies of the 4D theory,
based on Monte Carlo methods \cite{HMP,KS} found $\alpha_c = -8.2$
and $\alpha_c = -8.275$, respectively. A crude extrapolation of
the trend of our 4D results, as given in table
\ref{tabCrit2-4D_Phi4_jk}, gives a critical value of $\alpha_c
\approx -8.08$.

Remarkably, even in this model, the results using the linear
variational parameter only are no different. The critical mass
parameters $\alpha_c$, at all orders and in all dimensions are the
same. We also found that including the quartic parameter made no
difference to the critical behaviour.


\section{Exact Analysis of Ising Model case}

In this context, exact means no numerical evaluations were used.
The approximate algebraic forms produced for the free energy were
solved algebraically, including the optimisation aspect.  This is
possible only for Ising model where there is a single variational
parameter $j$.

\subsection{First order} \label{secforFtVP}

The lowest order, $R=1$, LDE approximation to the Ising model is
easily calculated. The derivation also provides a simple concrete
example of the LDE method. For the free energy we find that
\begin{eqnarray}
 f^{(1)} &=& - \ln \mathcal Z_0 - \cexpv {\Delta S}
 \label{f1ana1}
 \\
 \label{f1ana2}
 &=& - \ln \mathcal Z_0 - \kappa d \zexpv{\phi}^2
  - (j - J) \zexpv \phi
\end{eqnarray}
We are using statistical averages with respect to the $S_0$
action, defined in \tref{zexpdef} so that
\beq
  \label{f1ana3}
 Z_0 = I_0, \;\;\; \zexpv{\phi^p} = \frac{I_p}{I_0}
\eeq
They also factorise in the same way as the $\oexpv{Q}$ expectation
values \tref{1fact}. Note the absence of quadratic and quartic
terms from $\cexpv {\Delta S}$ because of the simple form of the
action and trial action for the Ising model. To find the minimum,
we need to differentiate with respect to the sole variational
parameter $j$, and set the resulting equation to zero. Using
\tref{dpzdj} we find that
\begin{equation} \label{f1ana5} \pderiv {f^{(1)}} {j} =
\left( 2 \kappa d \zexpv{\phi} + j - J \right) \left( \zexpv
{\phi^2} - \zexpv {\phi}^2 \right)
\end{equation}
We set the above equation to equal zero, and provided $\zexpv
{\phi^2} \neq \zexpv {\phi}^2$, the optimum $j$ value, $\jbar$, is
found by solving
\begin{equation} \label{f1ana6}
\jbar = J - 2 \kappa d \zexpv{\phi}
\end{equation}
The equation is generally  transcendental since $\zexpv \phi$
depends on $j$.

For the Ising model, using the form of the $I_p$ given in
\tref{isip}, it is possible to write this equation as
\begin{equation}
\label{is1jbar}
 \jbar  = J + 2 \kappa d \tanh (\jbar)
\end{equation}
In general this equation has three real solutions.  If we wish to
use the principle of minimum free energy we have to pick a
solution which actually produces a minimum.

However, one can quickly see that for $J=0$ then there is only one
solution in the region $\kappa\leq\kappa_c = 1/(2d)$ and that is
$j=0$. Further, $\partial^2 f^{(1)}/\partial j^2 =0$ at
$\kappa=\kappa_c$ where the $j=0$ solution turns from a global
minimum into a local maximum (as happens at higher odd orders, see
figure \ref{fisf7j}), and the correct value of $\jbar$ is a
non-zero one for $\kappa > \kappa_c$.

One can now see what the critical exponents are.  First let us
note that the delta expansion for $\expv{\phi}$ in the Ising model
gives to lowest order
\bea
 \expv{\phi} &=& \zexpv{\phi}
 + (j-J) \left(\zexpv{\phi^2}-(\zexpv{\phi})^2\right)
 +d \kappa \left(\zexpv{\phi^2}\zexpv{\phi}-(\zexpv{\phi})^3\right)
\\
 &=& -\tanh(j)
 + \left( (j-J)  +d \kappa \tanh(j) \right) (1-\tanh^2(j))
 \label{expphio1}
\eea
Near the critical point we expect $\jbar(J)$ to be small as $J \ra
0$, $\kappa \ra \kappa_c= 1/(2d)$. Normally we consider the
optimal variational parameter $j$ to be a function of the two
physical variables $\kappa$ and $J$.  To solve analytically,
however, its easiest if we express the relationship the other way
round, i.e.\ making expansions in terms of $\jbar$ and express the
one physical parameter we vary in any critical exponent expression
in terms of $\jbar$.  Thus keeping $\kappa=\kappa_c$ and studying
$J(\jbar)$, $f(J(\jbar))$, $\expv{\phi}(\jbar)$ in terms of a
small $\jbar$ expansion, we find that $J=\jbar^3/3 + \ldots$ and
$\expv{\phi} = -\jbar/2 + \ldots$ showing that the critical
exponent $\delta$ of \tref{deltaDef} is $3$. Likewise, for $J=0$
we find that $(\kappa/\kappa_c-1) = \jbar^2/3 + \ldots$,
$\expv{\phi} = -\jbar/2 + \ldots$ and so the exponent $\beta$ of
\tref{betadef} is $1/2$.

The exact same exponents are obtained in an analytical mean field
analysis.  So is the LDE using minimum free energy optimisation
the same as Mean Field theory? The answer is not exactly. In mean
field theory for the Ising model on a simple hypercubic lattice we
would write
\bea
Z &=& \int D\phi \exp \{ \kappa \sum_{i \in \Lambda} \sum_{j \in
\mathcal{N}_i^+}   \phi_i \phi_j
 - J \sum_{i \in \Lambda} \phi_i\}
 \\
\approx Z_\mathrm{MF} & :=& \int D\eta \exp \{ \kappa Nd v^2 - JNv
+ 2dv \kappa \sum_{i \in \Lambda}  \eta_i  - J \sum_{i \in
\Lambda} \eta_i\}
 \\
 &=& 2 Z_0 \cosh ( 2d\kappa v -J)
 \label{ZMF}
\eea
where $\eta_i=\phi_i-v$ and $v$ is defined, self-consistently, to
be the expectation value of the field, i.e.\
\bea
 v = \frac{1}{Z_\mathrm{MF}} \frac{\partial
 Z_\mathrm{MF}}{\partial J} = \tanh (2d\kappa v -J)
 \label{vmf}
\eea
Comparing the variational parameter here, $v$ with $j$ of the our
LDE calculation using lowest free energy optimisation,
\tref{is1jbar}, we see that the two are related by
\beq
\jbar  = J - 2d\kappa v
 \label{jbarvmf}
\eeq
However, the forms for the expectation values $\expv{\phi}$ in the
two calculations are clearly different as comparing
\tref{expphio1} and \tref{vmf}, using \tref{jbarvmf} clearly
shows. The difference though becomes negligible near the critical
point which explains why the two approaches give the same answer
for critical behaviour.

There is another way to view the link to mean field and that is to
realise that mean field is merely another optimisation scheme
within the LDE family.  If we chose our $S_0$ to be
\beq
 S_0 = \sum_{i \in \Lambda} L_0(\phi_i,\vecv), \;\;\;
 L_0(\phi,\vecv) = \Omega  + j \phi
 \label{LDEmf}
\eeq
rather than that of \tref{L0latdef} we see that the LDE parameter
$j$ is playing the role of a mean field, and comparing
\tref{LDEmf} with \tref{ZMF}, we obtain the relationship
\tref{jbarvmf} but for all $j$ values, not just the optimal
values.  If we used an optimisation scheme where
\beq
 \jbar:=  2d\kappa \expv{\phi} -J
\label{MFopt}
\eeq
then we would obtain MF as the zero-th order in this LDE
scheme.\tnote{It doesn't seem to work at next order.  Is there
another optimisation scheme that gives MF at lowest order and
works at higher orders?}

\subsection{Exact results for $\kappa_c$}\label{serkc}

The Ising model can be solved exactly within the LDE approximation
with minimum free energy optimisation at least at the orders and
for the dimensions studied here.  This was done using the
algebraic forms for the diagrams produced by our nauty and MAPLE
routines. Then by using the algebraic manipulation capabilities of
MAPLE, we searched for the point where $\partial^2
f^{(R)}/\partial j^2 =0$ at $j=0$ as numerically we know this is
where the critical point is.  The exact values are given in table
\ref{kapcjexact} as rationals, together with an eight digit
approximate decimal value.
\begin{table}[htb]
\begin{center}
\begin{tabular}{c|ccc}
 Order & \multicolumn{3}{c}{$\kappa_c$ for 2D} \\
 \hline \hline
1 & 1/4
\\
2 & 4/12 = & 1/3 = & .3333333333
\\
3 & 12/(104/3)  = & 9/26   = & .3461538462
\\
4 & (104/3)/92        = & 26/69   = & .3768115942
\\
5 & 92/(3608/15)         = & 345/902    = & .3824833703
\\
6 & (3608/15)/(9148/15)    = & 902/2287     = & .3944031482
\\
 7 & (9148/15)/(485788/315)     = & 48027/121447     = & .3954564543
\\
\hline
 & \multicolumn{3}{c}{$\kappa_c$ for 3D}  \\
\hline 1 &  1/6
\\
2 & 6/30 = & 1/5 = & .2000000000
  \\
3 & 30/148      = & 15/74  = & .2027027027
  \\
4 & 148/706           = & 74/353  = & .2096317280
  \\
5 & 706/(16804/5)        = & 1765/8402  = & .2100690312
  \\
6 & (16804/5)/15746        = & 8402/39365   = & .2134383335
  \\
 7 & 15746/(7742666/105)        = & 826665/3871333   = & .2135349762
   \\
\hline
 & \multicolumn{3}{c}{$\kappa_c$ for 4D}  \\
 \hline 1 & 1/8
\\
2 & 8/56 = & 1/7 = & .1428571429
\\
3 & 56/(1168/3) = & 21/146 = & .1438356164
\\
4 & (1168/3)/(7976/3) = & 146/997 = & .1464393180
\\
5 & (7976/3)/(272176/15) = & 4985/34022 = & .1465228382
\\
6 & (272176/15)/(614728/5) = & 34022/230523 = & .1475861411
\\
 7 & (614728/5)/(262409816/315) = & 4840983/32801227 = & .1475854242
\\
\hline
 & \multicolumn{3}{c}{$\kappa_c$ for d}  \\
 \hline 1 & $1/(2d)$
\\
2 & $(2d)/(4d^2-2d)$ = & $1/(2d-1)$
\\
3 & $(4d^2-2d)/(8d^3-8d^2+4d/3 (?) )$
\end{tabular}
\end{center}
\caption{Exact critical points for the 2D, 3D and 4D Ising models
The values are obtained by using the linear variational parameter
$j$ only. For 3D the Monte Carlo result is $\kappa_c = 0.221654$
\cite{HPV,FL,TB}.} \label{kapcjexact}
\end{table}
The simple values obtained reflect the simple nature of the model.
By way of comparison, in the spin-one model these fractions of
integers are replaced by ratios of polynomials in $\exp \{ -\alpha
\}$ where $\alpha$ is the extra physical parameter of the spin-one
model. The values of these integers for the Ising model results
depend on the details of the lattice used.  We have no explanation
for the fact that the same numbers appear twice in the first
column, once in the numerator, once in the denominator, though
related behaviour is known for the LDE expansion of the partition
function, at least in zero-dimensions (i.e.\ the simple integral
of $\exp \{ -x^2-\lambda x^4\}$ for $x$ from $-\infty$ to
$+\infty$).\tnote{The $\kappa_c$ values are calculated by using
only the linear variational parameter. As already mentioned,
although the quadratic term does not contribute to the full Ising
model, this is not necessarily the case for the expanded and
truncated model. We performed all the calculations using two
variational parameters, the linear $j$ and quadratic $k$, and
found no difference in the behaviour of the model.}

Note the absence of a 1D case in table \ref{tiskc}. This is
because the method does not produce any critical points in the 1D
case, which is physically correct --- there are no phase
transitions in the 1D Ising model. We have already alluded to
quantitative similarities between our model and mean field theory.
The fact that we do not find a phase transition in the 1D case is
a step better than mean field theory, which does predict a phase
transition in the 1D Ising model.

Finally, we also used MAPLE to check the critical behaviour in the
Ising model.  Using the expressions for three dimensions we were
able to show that the critical exponent $\delta$ is indeed
precisely $3$.

\section{Conclusions}\label{scon}

We have studied the LDE together with several different
optimisation schemes for the Ising, model, spin-one model and full
$\lambda \phi^4$ field theory.  In doing so we have gone four
orders higher than any previous LDE lattice study except for the
pure gauge study of Kerler and Metz \cite{KM}.  In any case, to
the best of our knowledge, the Ising and spin-one models have not
previously been studied on the lattice using LDE. To reach our
high orders we had to automate the whole diagrammatic expansion
procedure. While similar expansions (high-temperature etc.) have
been done to much higher orders elsewhere (e.g.\ see
\cite{HR99,CHPPV}), these non-LDE studies do not have the
optimisation stage of the LDE method.  This means the limitations
in terms of computing are quite different for LDE, we must
calculate the whole expression for the free energy many many more
times and our expressions have several extra parameters.  Thus we
have achieved one of our aims which was to demonstrate that high
orders of LDE lattice expansions can be calculated without using
significant computing resources.  In doing so we have indicated
some of the key issues and our solutions to several problems.

It is clear to us that the method could be improved in several
places and higher orders reached, e.g.\ through the use of the
`free diagrammatic expansion' \cite{ID}. Though we could never
match the orders produced for straight high-temperature
expansions, experience with LDE in QM has been that the
variational aspect more than compensates for the additional work
or in our case, low numbers of terms in our expansions.

We studied several different optimisation schemes, PMS (Principle
of Minimal Sensitivity) and FAC (Fastest Apparent Convergence) but
in this particular case, only our use of the minimum free energy
principle seemed to work.  It is hard to find detailed comparisons
of optimisation schemes or to find the use of the minimal free
energy condition in the literature of LDE on the lattice, so it is
hoped that our results have moved the debate on.

We also note that all our work, and that of the LDE lattice
literature, is exclusively on a simple cubic lattice, though of
various dimensions. It would be straightforward to repeat our
calculations for other lattices, just as high temperature
expansions have been done.  We merely note that calculations up to
third order on a hypertetrahedral lattice have shown the expected
variation in the position of critical points but there is no sign
of a change in the critical exponents.

We set out to use the calculation of critical exponents as a test
of the LDE method, at least for this lattice approximation.  As
such we have shown that the LDE method has failed to capture the
non-perturbative infra-red physics of scalar filed models near the
critical point as all our results have given the incorrect mean
field values. This is despite us pushing the calculation to far
higher orders and studying the region close to the transition
extremely carefully, taking great care to keep numerical errors
under control.  The fact that low orders produce these mean field
values is not too surprising, as the exact analysis of the Ising
model demonstrates, but we are unable to explain why the LDE
method does not start to move towards the correct values as we
study higher orders.  In most studies where exact results are
known (typically quantum mechanics) moving to higher and higher
orders in the LDE method rapidly brings one to good approximations
of non-perturbative results.  Our conclusion is that this form of
LDE on the lattice is unable to access the non-perturbative
aspects of field theories near the critical point.

This is not necessarily the end of the use of LDE for field
theories.  We are studying a theory which is not-asymptotically
free, using a space-time lattice and only looking at one type of
physics.  LDE calculations for QCD, using continuous space-time
methods\footnote{We note, however, that that the results of Gandhi
and McKane \cite{GM} for critical exponents using LDE with
continuum space-time methods were only slightly better.  See also
the work of Ogure and Sato \cite{OS98a,OS99}, and Chiku
\cite{Chi00}.} or even for other quantities might show better
results. One might try to find a different way of applying the LDE
in this case. After all, the high-temperature expansion gives a
series which itself necessarily has \emph{no} transitions yet by
using Pad\'e methods and similar one can extract remarkably good
results.  Such methods are not directly applicable to LDE as it
does \emph{not} calculate coefficients of a series.  LDE only
provides a sequence of results which in some testable cases
converge extremely fast to the correct answer. For the LDE the
most obvious variation is to try different ansatz for the trial
action $S_0$. We have indicated that it must be ultra-local for
calculational reasons, but beyond that we have great flexibility.
For instance we tried ansatz such as\tnote{Exactly what did we
try? Surely best if $k=0$ ansatz is used.}
\beq
 L_0 = j \phi + k \phi^2 + q |\phi|^p + l \phi^4
\eeq
with non-polynomial power $p$ a variational parameter.  However
our studies at low orders showed no promise and we did not pursue
this further.

\begin{acknowledgements}
We would like to acknowledge S.Hands, H.F.Jones, R.J.Rivers and
A.Ritz for many useful comments.  TSE would like to thank
K.George, R.Johnson, K.Kiyani, C.Mukherjee, R.Richardson and
V.Stojevic for their contributions to the preliminary
investigations of these models.
\end{acknowledgements}


\appendix

\section{Growing unique configurations}\label{aguc}

For diagrams of $n$ links (an abstract graph), we could easily
construct all possible configurations (embeddings of this graph on
a space-time lattice) by throwing down $n$ links in all possible
ways on all the lattice graph edges in an $(n+1)^d$ size box, but
every configuration then appears in $(n+1)^d$ copies differing
only by translation invariance.  This is clearly highly
inefficient at the orders and dimensions we work at so the aim of
the algorithm outlined here is to produce a complete set of
configurations of a certain size, once and only once, modulo
translation symmetry. The algorithm is inspired by those described
for percolation simulations \cite{SA,Me90}.

The starting point is to define some relation on the edges in the
configurations, so that we can always decide that any certain line
is `greater' or `less' than any other line. To do this, we first
have to define a relation on the points of the lattice, and so we
say that a point $\xvec=(x_1,x_2,\ldots,x_d)$ is \emph{greater}
than a point $\yvec$, i.e.\ $\xvec > \yvec$,  if the first
coordinate of $\xvec$ is greater than the first coordinate of
$\yvec$, $x_1>y_1$. If the two coordinates happen to be equal,
then we compare the second coordinates in the same way. If these
turn out to be equal, we continue moving on to the next coordinate
until we have exhausted all the $d$ coordinates that label a point
on the lattice.
\beq
\xvec > \yvec \Leftrightarrow \exists \; i \; \mathrm{s.t.} \; x_i
> y_i \mbox{ and } x_j = y_j \; \forall \; j<i, \;\;\; i,j\in
\{1,2,\ldots, d\}
\eeq
Of course, if all the coordinates turn out to be the same, the two
points $p_1$ and $p_2$ are equal.

In our configurations edges, have no directions but it is always
useful to list them with the least vertex first. An edge is then
an ordered pair of two (neighbouring) points $\eline =
(\xvec,\xvec_\mathrm{nn}), \xvec < \xvec_\mathrm{nn}$. To define a
relation between two edges $\eline$ and $\fline$, we say that
$\eline$ is \emph{greater} than $\fline$ if $\eline$'s first
vertex is greater than the first vertex of $\fline$. If the two
points happen to be equal, then we compare the second coordinate
in the same way.
\beq
\eline > \fline \Leftrightarrow \exists \; i \; \mathrm{s.t.} \;
e_i
> f_j \mbox{ and } e_j = f_j \; \forall \; j<i, \;\;\;
i,j\in \{1,2\}
 \label{linerel}
\eeq

The relation we defined on the lines will ensure that \emph{every
configuration} has a \emph{unique least} line.  This link goes
through the least vertex of the configuration which we choose to
be the origin $\mathcal{O}$ on the space-time lattice.   All the
configurations we create have the origin as the least vertex as
then we are guaranteed that all our configurations are not related
by translation symmetry to any of the others we produce.

What then remains to be done is to ensure that we grow \emph{all
possible} configurations \emph{once and only once} from this
lowest vertex origin. A recursive algorithm which does this is as
follows. For an existing configuration of $n$ links, $C_n$, the
set of possible lines to be added is an ordered list of links,
$\calL_n$, in which each link satisfies the following five
conditions:
\begin{enumerate}
\item they never contains a vertex less than the origin
$\mathcal{O}$ (eliminates translations),
\item they start from an existing vertex in the configuration
(we are not interested in disconnected graphs),
\item they do not lie on top of any other links (we are not interested
in non-simple graphs),
\item they are always greater than the least link in the existing
configuration $C_n$ (ensures we visit each configuration only
once, and none are related by translation invariance as the first
and least line remains the least line)
\item each link has not been investigated at previous orders even
if it is not in the existing configuration $C_n$ (ensures each
configuration is visited once and only once)
\end{enumerate}
We then add a link to $C_n$, starting with the least link in the
list of possible links $\calL_n$, creating a configuration
$C_{n+1}$ of $n+1$ links. We then recurse, considering possible
links to add to this $n+1$ link configuration, $C_{n+1}$.  We stop
recursing when $R$ links have been added. Once we have considered
adding one link to the $n$-link configuration, say $e_i \in
\calL_n$, and all the possible larger configurations coming from
that, before we consider the next possible link, $e_{i+1} \in
\calL_n$ for this n-link configuration $C_n$, we mark the link
just investigated, $e_i$,  in the lattice as having been
\emph{used}. This enables the implementation of the fifth
condition in the production of the lists $\calL_m$ ($m>n$) and is
vital to eliminate repetitions of the same configuration.

Note we do not generate configurations in the order of their
links. The relation \tref{linerel} gives a natural order for the
links in every configuration but many configurations have links
which are connected to each other only via links which are greater
than both of them.  This is why the fourth condition only requires
potential new links to be greater than the first one.

To see how this works, or to test the voracity of alternative
algorithms, the example of $d=2$ space-time dimensions on a square
lattice up to configurations of $R=3$ links is often sufficient.
For instance consider the two size 2 configurations of figure
\ref{f3poss}.
\begin{figure}[!htb]
\begin{center}
\includegraphics{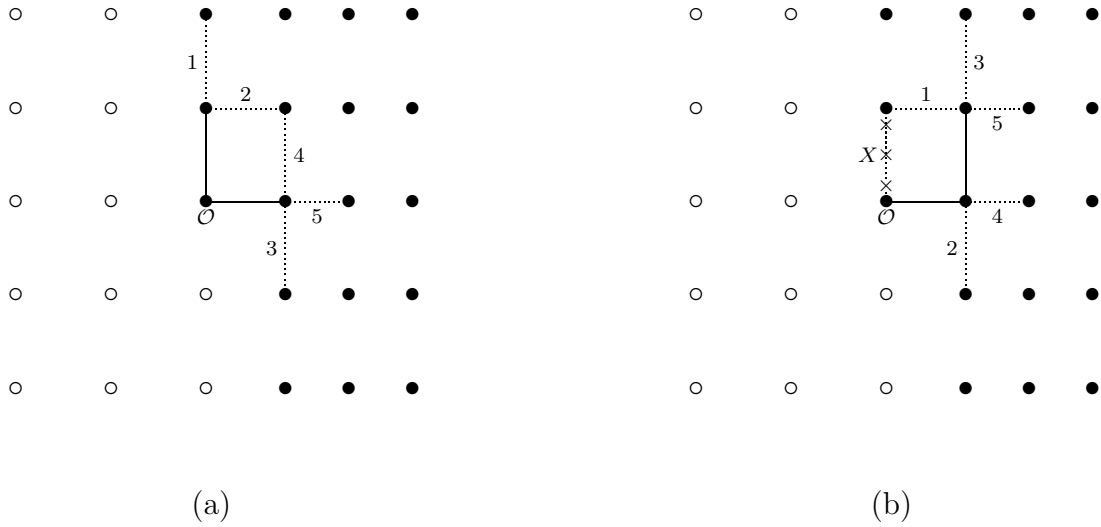}
\end{center}
\caption{Two different second order configurations are shown in
  solid lines.  The first coordinate is measured up from the
  origin $\mathcal{O}$, the second coordinate is measured to the
  right of the origin.  Thus the vertex points with open circles
  are never used to ensure that the
  origin $\mathcal{O}$ is always the smallest vertex, thus
  eliminating space-time translations of diagrams.
The links which can be added to produce third order diagrams are
given as dashed lines, numbering reflecting the edge ordering as
defined in the text, i.e.\ $e_i < e_j$ if $i<j$. Edge X in case
(b) is crossed out to indicate it is excluded by our rule five
ensuring that when link 4 is added to configuration (a), the
resulting order 3 configuration is not produced from configuration
(b).} \label{f3poss}
\end{figure}
With the first coordinate up the page, the second to the right of
the origin $\mathcal{O}$, we exclude the lattice vertices shown as
open circles by rule one and so eliminate repetitions related by
space-time translation.  It also follows that configuration (a) is
generated before configuration (b) so this is why the link
labelled X has can not be used in the latter, by rule five.
Without this limitation, the order 3 configuration using link 4 in
case (a) would have been reproduced by case (b) with link X since
the latter link is legitimate by all the other rules.

\section{Free Energy Density}
\label{secnumFED}

The calculation of the free energy density is based on the
formalism developed in section \ref{secforCtFED} with the free
energy density given as an expansion of order $R$
\begin{equation} \label{eqfROrder}
f^{(R)}
 = - \sum_{n = 0}^R \frac{\delta_2^n}{n!} f_{1n}
 = - \sum_{n = 0}^R \frac{\delta^n}{n!} f_n
\end{equation}
We also know that $f_0 = \ln \mathcal{Z}_0$, and for the other
$f_n$'s, we have to make use of all the different connected graphs
of size $n$ and their multiplicities. For $n = 1$ this is a
particularly easy task, but the $n = 2$ term is illustrative, yet
simple enough to work through as an example.

The diagram counting is the first step in the numerical evaluation
of any $f_n$. The \verb|C|/\verb|C++| program implementing the
algorithm for counting multigraphs without loops is used to find
the multiplicities of all the graphs of size 2 and less. Note that
the multiplicities will depend on the dimension of the problem,
i.e.\ the dimension of the lattice that the configurations are
generated on. For definiteness we will consider the example of a
$d = 3$ cubic lattice. The program will output the results in a
text file, and for our example we get
\begin{equation} \label{eq2Mapgraphs1}
3 \ \
 \setlength{\unitlength}{1pt}
 \begin{picture}(20,10)
 \put(0,2){\circle*{4}}
 \put(0,2){\line(1,0){20}}
 \put(20,2){\circle*{4}}
 \end{picture}
 \hspace{2cm}
 3 \ \
 \setlength{\unitlength}{1pt}
 \begin{picture}(20,10)
 \put(0,2){\circle*{4}}
 \put(0,0){\line(1,0){20}}
 \put(0,4){\line(1,0){20}}
 \put(20,2){\circle*{4}}
 \end{picture}
 \hspace{2cm}
 30 \ \
 \setlength{\unitlength}{1pt}
 \begin{picture}(40,10)
 \put(0,2){\circle*{4}}
 \put(0,2){\line(1,0){20}}
 \put(20,2){\circle*{4}}
 \put(20,2){\circle*{4}}
 \put(20,2){\line(1,0){20}}
 \put(40,2){\circle*{4}}
 \end{picture}
\end{equation}
The numbers preceding the graphs are the multiplicities, the
number of distinct ways these graphs can be embedded on a
space-time lattice up to translations. Thus $f_{11},f_{22}$ are
\bea
 f_{11} &=&
  3 \cexpv{ \
 \setlength{\unitlength}{1pt}
 \begin{picture}(20,10)
 \put(0,2){\circle*{4}}
 \put(0,2){\line(1,0){20}}
 \put(20,2){\circle*{4}}
 \end{picture}
 \ }
  \label{ef11ex}
\\
 f_{12} &=&
  30 \cexpv{ \
 \begin{picture}(40,10)
 \put(0,2){\circle*{4}}
 \put(0,2){\line(1,0){20}}
 \put(20,2){\circle*{4}}
 \put(20,2){\circle*{4}}
 \put(20,2){\line(1,0){20}}
 \put(40,2){\circle*{4}}
 \end{picture}
   \
   }
 + 3 \cexpv {\
 \begin{picture}(20,10)
 \put(0,2){\circle*{4}}
 \put(0,0){\line(1,0){20}}
 \put(0,4){\line(1,0){20}}
 \put(20,2){\circle*{4}}
 \end{picture}
 \ }
\eea
The cumulant expectation values then have to be expanded in terms
of statistical averages using equation \eqref{GenCumDef}. For
example,
\begin{equation} \label{Cum2Decomp}
\cexpv {\
 \begin{picture}(40,10)
 \put(0,2){\circle*{4}}
 \put(0,2){\line(1,0){20}}
 \put(20,2){\circle*{4}}
 \put(20,2){\circle*{4}}
 \put(20,2){\line(1,0){20}}
 \put(40,2){\circle*{4}}
 \end{picture}
 \ }
 = \oexpv {\
  \begin{picture}(40,10)
 \put(0,2){\circle*{4}}
 \put(0,2){\line(1,0){20}}
 \put(20,2){\circle*{4}}
 \put(20,2){\circle*{4}}
 \put(20,2){\line(1,0){20}}
 \put(40,2){\circle*{4}}
 \end{picture}
 \
 }
  - \oexpv {
  \
  \begin{picture}(20,10)
 \put(0,2){\circle*{4}}
 \put(0,2){\line(1,0){20}}
 \put(20,2){\circle*{4}}
 \end{picture}
 \
 }
 \oexpv {
 \
 \begin{picture}(20,10)
 \put(0,2){\circle*{4}}
 \put(0,2){\line(1,0){20}}
 \put(20,2){\circle*{4}}
 \end{picture}
 \
 }
\end{equation}
These graphs can be in terms of fields, physical and variational
parameters, and the expectation values with respect to the $L_1$
trial Lagrangian fields at one site. Using \tref{Statfact} these
last contributions are expressed in terms of the $J_p$ integrals
of \tref{Jndef}.  For instance
\beq \oexpv {\
  \begin{picture}(40,10)
 \put(0,2){\circle*{4}}
 \put(0,2){\line(1,0){20}}
 \put(20,2){\circle*{4}}
 \put(20,2){\circle*{4}}
 \put(20,2){\line(1,0){20}}
 \put(40,2){\circle*{4}}
 \end{picture}
 \
 }
 =
\frac{J_1}{J_0} \frac{ J_1}{J_0}
 \label{Cum2Fields}
\eeq
However, the $J_n$ integrals contain $\delta_1$ in their exponents
through $L_1$. Terms such as $f_{12}$ are already order
$(\delta_2)^2$ so they can have their $J_n$ integrals simply
replaced by $I_n$ of \tref{Ipdef}. However $f_{11}$ have only one
power of delta so they require us to expand the $J_n$ integrals to
first order in $\delta_1$ and the terms from $f_{10}=\ln(J_0)$
require full second order expansions. For the Ising model case we
have
\beq
 J_n := I_n + \delta_2\Delta J I_{n+1} + \half(\delta_2\Delta J)^2I_{n+2}
 + O((\delta_2)^3)
\eeq
where $\Delta J = J-j$. Inserting into \tref{ef11ex} truncating at
order of delta $R=2$ and then setting the $\delta$'s equal to one,
we finally get the following expression for second order term of
the full free energy, $f_2$:
\bea
f_2 & =& 30 \left[ \zexpv {\phi}^2 \zexpv {\phi^2} - \zexpv
{\phi}^4 \right] + 3 \left[ \zexpv {\phi^2}^2 - \zexpv {\phi}^4
\right]
 \nonumber \\
&& + 12 \Big[ \zexpv {\phi} \left[ (j - J) \zexpv {\phi^2} + (k -
\alpha)
  \zexpv {\phi^3} + (l - g) \zexpv {\phi^5}  \right]
 \nonumber \\
&& - \zexpv {\phi}^2 \left[ (j - J) \zexpv {\phi} + (k - \alpha)
\zexpv {\phi^2} + (l - g) \zexpv {\phi^4}  \right] \Big]
 \nonumber \\
&& + (j-J)^2 \zexpv {\phi^2} + (k-\alpha)^2 \zexpv {\phi^4} +
(l-g)^2 \zexpv {\phi^8}
 \nonumber \\
&& + 2 \left[ (j-J) (k-\alpha) \zexpv {\phi^3} + (j-J) (l-g)
\zexpv {\phi^5} + (k-\alpha) (l-g) \zexpv {\phi^6} \right]
 \nonumber \\
&& - (j-J)^2 \zexpv {\phi}^2 - (k-\alpha)^2 \zexpv {\phi^2}^2 -
(l-g)^2 \zexpv {\phi^4}^2
 \nonumber \\
 && - 2 \big[ (j-J) (k-\alpha) \zexpv {\phi} \zexpv {\phi^2} +
(j-J) (l-g) \zexpv {\phi} \zexpv {\phi^4}
 \nonumber \\
 &&+ (k-\alpha) (l-g) \zexpv {\phi^2} \zexpv {\phi^4} \big]
 \label{eqf2numfinal}
\eea
where $\zexpv {\phi^p} = I_p/I_0$.

We have shown the above expression in its entirety to demonstrate
the general form of the $f_n$. Thus we see that the calculation of
$f_n$ is broken down into a sum of terms (the number of which will
get rather large as we go to higher orders) depending on the
physical parameters $J$, $\alpha$ and $g$, the variational
parameters $j$, $k$ and $l$, and also the field expectation values
$\zexpv {\phi^p}$. Note that the $\zexpv {\phi^p}$ actually depend
\emph{only} on the variational parameters. From a numerical
perspective, however, we look upon the $\zexpv {\phi^p}$ as
quantities in their own right since their evaluation is a
non-trivial task. Thus, mathematically we write $f^{(R)} (J,
\alpha, g; j, k, l)$ to indicate the variables which contribute to
the calculation of $f^{(R)}$. On the other hand, numerically we
recognise that in order to calculate $f^{(R)}$, we also need to
address separately the calculation of a set of $\zexpv
{\phi^p}$'s.

Maple is the ideal tool to perform all the necessary expansions,
starting from the appropriate list of diagrams and multiplicities
such as \tref{eq2Mapgraphs1}, through to obtaining the expressions
for the $f_n$. For the particular case of $n = 2$, this will be an
expression of the form shown in equation \eqref{eqf2numfinal}
above. Maple will output $f_2$ as code in the \verb|C| programming
language, optimising the sum for minimum computation. We note that
this output was always checked up to 3rd order against code that
we had available from the study of a complex $\phi^4$ theory
\cite{EIM}. By performing this 3rd order check routinely, we
gained confidence in the validity of our diagram counting and code
generating techniques.

\section{Numerical Accuracy} \label{secnumNA}

The standard double precision available to \verb|C|/\verb|C++|
(type \verb|double|) is limited to around 17 decimal places of
precision on floating point numbers.\footnote{Technically, this
will depend on the computer architecture and the particular
\textrm{C/C++} implementation. We used a standard PC based on an
AMD Athlon XP 2100+ processor (a 32-bit architecture), and we used
Linux with the GNU Compiler Collection v3.2 implementation of
\textrm{C/C++}.} The nature of the calculations in our model is
such that there are many additions and subtractions, with a very
wide range of values. Here we have to deal with the common problem
of terms in a summation conspiring to cancel out in such a way
that smaller terms fail to contribute to a sum in which their
contribution is, in fact, important. This is the \emph{rounding}
problem.

\subsection{Free Energy Density}

Now consider an example from our actual model --- the calculation
of $f^{(R)}$ for a set of physical and variational parameters. The
numerical values of the parameters in this example are taken from
one of the points visited by the program as it searches for the
critical point. This is a realistic scenario for the double
precision program, since it is taken at a stage where the program
is exploring variations of the order of $10^{-15}$ in the free
energy. This is the sort of accuracy we would expect to achieve
from an implementation using \verb|C|/\verb|C++|'s standard double
precision. The values of interest are listed in table
\ref{tabFreeNums}.
\begin{table}[!htb]
\begin{center}
\begin{tabular}{|c|c|c|c|}
\hline
$f_n$ & Smallest & Largest & Total \\
\hline $n = 1$ & $1.4997840002 \cdot 10^{-10}$ & $2.7472168610
\cdot
10^{-10}$ & $-9.1453475295 \cdot 10^{-11}$ \\
$n = 2$ & $4.6741039631 \cdot 10^{-22}$ & $7.4509483795 \cdot
10^{-2}$
& $-7.4509483756 \cdot 10^{-2}$ \\
$n = 3$ & $2.0210522783 \cdot 10^{-29}$ & $7.4509483795 \cdot
10^{-2}$
& $-6.4570642135 \cdot 10^{-11}$ \\
\hline
\end{tabular}
\end{center}
\caption{The table summarises the absolute values of the smallest
and largest terms in the sums which constitute the $f_n$ terms,
for all orders up to the third. The total value of each order is
given too. These particular values are taken from a real
calculation of $f^{(R)}$ close to the critical point. }
\label{tabFreeNums}
\end{table}

The third order proves already to be numerically both interesting
and challenging. The smallest term being of the order of
$10^{-29}$ and the largest of $10^{-2}$. The total result is of
the order of $10^{-11}$, and is arrived at by summing 93
terms\footnote{In the 4D case.} whose values will range from the
smallest to the largest quoted in the table. It is obvious that if
we are limited to double precision floating point arithmetic, we
could easily end up completely losing contributions from terms
that might actually be important, especially, one suspects, near
physical transition points. One could argue that since the largest
term is of the order of $10^{-2}$ and the total is about
$10^{-11}$, we are actually only in danger of losing approximately
9 decimal places in this case, which should still be well covered
by the 17 decimal places that double precision offers.

However it is not the free energy itself but differences in the
free energies for similar parameter values which we are interested
in, as we are trying to find a set of variational parameters which
minimise the free energy.  For instance, the values discussed
above are derived from a point which is used to calculate
differences in free energies of size $10^{-15}$ only. Requiring
greater accuracy of the free energy would require calculations
involving even smaller values of the variational parameters, which
also means a greater range of values appearing in the terms that
contribute to the $f_n$.

Even if this level of accuracy is acceptable for this 3rd order
case that we are presenting, higher orders will further increase
the disparity of the terms. For all these reasons we decided to
use an arbitrary precision package, the GNU Multiple Precision
(GMP) \cite{GMP} library.  GMP fixes a number of bits which are
allocated to the mantissa of a floating point number. This means
that by setting the precision of the mantissa to, for example, 128
bits, the largest number we can accurately represent is
approximately $2^{128} \approx 3.4 \cdot 10^{38}$. In other words,
we will have approximately 38 decimal places of accuracy. When we
discuss the results we will find numerous occasions where we
really use the extra precision made available to us by the GMP
library. For instance, we were able to minimise the free energy
density to one part in $10^{-40}$.

\subsection{Integration} \label{secNumAccInt}\label{secnumI}

It has been emphasised that the calculation of the free energy
density (and for that matter, any other quantity with the LDE
formalism) will depend in particular on field expectation values
at a single site taken with respect to the trial action $\mathcal
S_0$ which depends on the variational parameters $j$, $k$ and $l$,
and of the form
\begin{equation}
\mathcal S_0 (j, k, l) = j \phi + k \phi^2 + l \phi^4
\end{equation}
We introduced in \tref{Ipdef} the notation $I_p$ for the general
form of the relevant integrals
\begin{equation} \label{eqIpee}
I_p = \intall \diff \phi \, \phi^p \exp \left[ - j \phi - k \phi^2
- l
  \phi^4 \right]
\end{equation}
These are trivial for the Ising and spin-one models but
non-trivial for the full $\phi^4$ model.  This is the case we
consider in this section.

Being quartic in the exponential, the calculation of the $I_p$'s
is a necessary numerical consideration. Numerical integration
comes with its own set of problems though. For small $j$, the most
interesting case for transitions, and for $p$ odd, the integrand
is nearly odd so to avoid severe rounding problems we used the
half-range integral forms
\bea \label{HalfInts}
I_p &=& 2 \intpos \diff \phi \, \phi^p \text{cosh} \left(j \phi
\right) e^{-k \phi^2 - l \phi^4} \qquad \text{for} \ p \
\text{even}
\\
I_p &=& 2 \intpos \diff \phi \, \phi^p \text{sinh} \left(-j \phi
\right) e^{-k \phi^2 - l \phi^4} \qquad \text{for} \ p \
\text{odd}
\eea
Having reduced the $I_p$'s to integrals over half the real line,
we have (approximately) halved the number of interpolations that
need to be performed in order to numerically evaluate the
integrals. This will be an obvious computing time bonus, but we
have also given ourselves a break from the dangerous
`rounding-off' error that appears in sums when terms are of
opposite signs.

Another important issue is the infinite range of integration. Our
integrand is suppressed at large $\phi$ by the negative terms in
the exponential. We will typically have $l = 25$, and, for
example, if we were to introduce the cutoff for the upper limit at
$\phi = 3$, the exponential would look like $\exp \{- 3j - 9k -
2025\}$.  Thus the variational parameters $j$ and $k$ would have
to be significantly large negative numbers to create a dent in the
assumption that this exponential is vanishingly small, and in
practice they are of the same order of magnitude as the
corresponding physical parameters and so $O(10)$ at most. So it is
easy to find a suitable cutoff value for $\phi$ in the $I_p$
integrals.  The numerical tests discussed below make it clear that
we can indeed trust our numerical integration.

Note that for \emph{every single} evaluation of the free energy in
the $\phi^4$ model, which will depend on the physical parameters
as well as the variational parameters, we will have to evaluate a
\emph{complete} set of $I_p$'s, from $p = 0$ through to whatever
the upper limit on $p$ turns out to be. For example, in a 3rd
order expansion we are  looking at the range $0 \le p \le 12$, the
upper limit coming from a term in which the quartic $\phi^4$ from
the exponential is cubed upon expansion. For a 7th order expansion
this range will thus go up to 28. We used a standard Romberg
integration method, with additional polynomial extrapolation of
successive refinements to zero stepsize \cite{NR}. However we
adapted this to perform the integration of all the required
$I_p$'s in parallel. Thus, instead of working with a single
integrand (corresponding to a specific value of $p$), we
manipulate a whole vector of elements, one entry for the integrand
at each value of $p$ required.

The integrals and the numerical methods used to calculate them are
not unusual.  However, we did require much higher accuracy than
normal so we will note how we tested our integration. For the
purpose of testing, we used the following values for the
variational parameters:\footnote{The apparent randomness of the
values is due to the fact that these constituted a point of
contention in the early stages of development of the program; it
subsequently assumed a status of a testing point.}
\begin{subequations} \label{eqVals}
\begin{align}
j &= 8.16343354579642367932991357923 \cdot 10^{-10} \\
k &= -7.6 - 1.58180882338368170019383733935 \cdot 10^{-9} \\
l &= 25
\end{align}
\end{subequations}
Using 256 bit floating point arithmetic, approximately 77 decimal
places of precision, table \ref{tabInts} shows a comparison of our
results for the above integrals for various values of $p$ with the
best results we could get for the same integral calculated using
numerical integration in Mathematica
\cite{Mathematica}.\tnote{Note that we could not get Mathematica
to give us results accurate to more than 50 decimal digits.}
\begin{table}[!htb]
\begin{center}
\begin{tabular}{|c|l|}
\hline
$p$ & \multicolumn{1}{|c|}{$I_p$} \\
\hline \hline $0$ &
$1.67769083077124890028663394770701193593$ \\
&
$1.67769083077124890028663394770701193592521428271904249117805060996740$
\\
\hline $1$ &
$1.81907694824780353539804027376 \cdot 10^{-10}$ \\
&
$1.81907694824780353539804027376452421473264747834415743104674406673865
\cdot 10^{-10}$ \\
\hline $6$ &
$0.0143833783157673970638326143289412430245$ \\
&
$0.01438337831576739706383261432894124302449287542710830784439069879080$
\\
\hline $11$ &
$5.606129663619230401281915955874438188097 \cdot 10^{-13}$ \\
&
$5.60612966361923040128191595587443818807152405548720830361149104778581
\cdot 10^{-13}$
\\
\hline $12$ &
$0.00068673673058880728698436218620528130020$ \\
&
$0.00068673673058880728698436218620528130019933560098958463918835374638$
\\
\hline
\end{tabular}
\end{center}
\caption{The table shows results of evaluating the integrals $I_p$
for various $p$, with the variational parameters as given in
\eqref{eqVals}. For each value of $p$ we show the result from
Mathematica (first row), and the result from our integration
routine (second row). Our results are calculated with 256 bit
floating point arithmetic; we show approximately 70 decimal places
(we expect the results to be that accurate).} \label{tabInts}
\end{table}

A slightly different check was performed on the $I_0$ integral for
$j=k=0$ and $l=1$ and $l=25$, which is simply
$2\Gamma(5/4)(l)^{-1/4}$. Mathematica has no problem evaluating
the Gamma function to very high precision. We found that our
integration routine matched the 150 decimal places of
Mathematica's results.\footnote{To achieve this accuracy we used
512 bit floating point arithmetic.}

\section{Benefits of High Precision Arithmetic}
\label{abhpa}

Many studies based on LDE compatible methods (e.g.\
\cite{GM,OS98a,OS99,Chiku00}) noted the similarity between their
critical exponents and those obtained by mean field theory.
Indeed, in preparing the results for \cite{EIM}, we also noticed
that critical exponents calculated with our 3rd order LDE method
produce values equivalent to mean field theory results. For this
reason, one of the main goals of this study was to estimate
critical exponents with high numerical accuracy, in particular to
establish any possible deviations from the mean field results as
higher orders of the LDE are considered.

As a prerequisite, we must find the critical point. Table
\ref{tabresIncAcc} summarises the estimates of $\kappa_c$ for the
3D Ising model at 3rd order. The values were found by searching
for the value of $\kappa$ where $\expv {\phibar}^{(3)} (\kappa, J
= 0)$ switches between zero and non-zero values.
\begin{table}
\begin{center}
\begin{tabular}{|c|l|}
\hline
Arithmetic / Tolerance & \multicolumn{1}{c}{$\kappa_c$} \vline \\
\hline \hline
64 bit /  $10^{-10}$ & 0.2027041103 \\
\hline
64 bit /  $10^{-15}$ & 0.202702704077629 \\
\hline
128 bit / $10^{-10}$ & 0.2027041103 \\
\hline
128 bit / $10^{-15}$ & 0.202702704077255 \\
\hline 128 bit / $10^{-20}$ &
0.20270270272418008017 \\
\hline 128 bit / $10^{-25}$ &
0.2027027027027865987084316 \\
\hline 128 bit / $10^{-30}$ &
0.202702702702702784632396012135 \\
\hline 256 bit / $10^{-20}$ &
0.20270270272418008017 \\
\hline 256 bit / $10^{-25}$ &
0.2027027027027865987084316 \\
\hline 256 bit / $10^{-30}$ &
0.202702702702702784632395797442 \\
\hline 256 bit / $10^{-35}$ &
0.20270270270270270302274056635402966 \\
\hline 256 bit / $10^{-40}$ &
0.2027027027027027027039528506075906985909 \\
\hline
\end{tabular}
\end{center}
\caption{The table summarises the critical inverse temperatures
$\kappa_c$ found for the 3D Ising model at 3rd order. Each row
represents a run with a different precision on the internal
arithmetic and tolerance value for the simplex optimisation. The
values of $\kappa_c$ are given to as many decimal places as is
allowed by the tolerance required on the minimum of $f^{(3)}$. }
\label{tabresIncAcc}
\end{table}
We will now explain the significance of the results given in the
table.

The results are all obtained with an implementation using the GNU
multiple precision library (GMP). The first two rows show results
obtained using 64 bits of precision, which is comparable to the
built-in double precision arithmetic of \verb|C/C++|. The
remaining rows were calculated with significantly higher
precision. 128 bit arithmetic gives us approximately 38 digits of
precision, while for 256 bit arithmetic we get approximately 77
digits. Every row is also labelled by a certain tolerance, which
is the termination criterion for the simplex minimisation routine.

To understand what the numbers are telling us, we have to explain
the route by which they are obtained. The quantity which is
minimised to the tolerance setting is the third-order free energy
density $f^{(3)}$ of the free energy. The $I_p$'s are evaluated to
high accuracy since they are simple functions in the Ising model.
Once evaluated, these are combined with the physical and
variational parameters to form $f^{(R)}$. The minimisation routine
keeps the physical parameters ($\kappa$ and $J$) constant, but
varies the variational parameters (only $j$ in this case) until it
hits a minimum in $f^{(R)}$. The procedure will continue
minimisation until it is satisfied that it has reached the
tolerance which is required of it. That is why we quote $\kappa_c$
in table \ref{tabresIncAcc} with as many decimal places as the
free energy was minimised to.

Although $f^{(R)}$ is minimised to within some quoted tolerance of
the minimum point, there is \emph{no reason to assume} that the
variational parameters have been fixed to the same accuracy. In
fact, experience shows that relatively large variations in the
variational parameters (of the order of $\epsilon$, say) produce
relatively small variations in $f^{(R)}$ (of the order of
$\epsilon^2$). This should come as no surprise, because we expect
linear variations not to contribute at a stationary point.
Moreover, this lends support to the reasoning of the PMS
criterion, since it is our goal to fix the variational parameters
in a way which will leave the physics as independent of the
variational parameters as possible.

For this simple Ising model case the exact location of the
critical point is $15/74 \approx 0.2027027027\ldots$ (see section
\ref{serkc}). We see in table \ref{tabresIncAcc} that the
numerical accuracy in $\kappa_c$ increases as we increase the
tolerance as long as this is paired with enough accuracy provided
by the internal arithmetic. If we require a tolerance of
$10^{-40}$ on the free energy, we are confident that $j$ is
correct to within approximately $10^{-20}$ of its value.  This
explains why the values of $\kappa_c$ in table \ref{tabresIncAcc}
follow the exact result up to about half of the decimal places
displayed.


\begin{thebibliography}{99}

\bibitem{MM}
 I.Montvay and G.M\"unster,
 \tbktitle{Quantum Fields on a Lattice},
 Cambridge University Press, 1994.

\bibitem{ZJ}
 J.Zinn-Justin,
 \tbktitle{Quantum Field Theory and Critical Phenomena},
 (Oxford University Press, Oxford, 1996, 3rd edition).

\bibitem{DM88}
 A.Duncan and M.Moshe,
 \tarttitle{Nonperturbative physics from interpolating actions}
 Phys.Lett. \vol{215} (1988) 352.


\bibitem{Jo}
 H.F.Jones,
 Nucl.Phys.B (Proc.Suppl.) \vol{39} (1995) 220.


\bibitem{Betal87}
 C.M.Bender et al.,
 \tarttitle{Logarithmic
  Approximations to Polynomial Lagrangeans}
  Phys.\ Rev.\ Lett.\ \textbf{58} (1987) 2615.

\bibitem{Betal88}
  C.M.Bender et al.,
  \tarttitle{Novel
  perturbative scheme in quantum field theory}
  Phys.\ Rev.\ D \textbf{37} (1988) 1472.

\bibitem{BJ88a}
  C.M.Bender and H.F.Jones,
  \tarttitle{New
  nonperturbative calculation: Renormalization and the triviality of
  $(\lambda \phi^4)_4$ field theory}
  Phys.\ Rev.\ D \textbf{38} (1988) 2526.

\bibitem{BJ88b}
 C.M.\ Bender and H.F.\ Jones,
 \tarttitle{Evaluation of
  Feynman diagrams in the logarithmic approach to quantum field
  theory}
  J.\ Math.\ Phys.\ \textbf{29} (1988) 2659.


\bibitem{St81}
 P.M.Stevenson,
 \tarttitle{Optimised Perturbation Theory}
 Phys.Rev.\vol{D23} (1981) 2916.

\bibitem{CH98}
 S.Chiku and T.Hatsuda,
 \tarttitle{Optimised perturbation Theory at Finite Temperature}
 PRD \vol{58} \yr{1998} 076001
 [\eprint{hep-ph/9803226}].

\bibitem{St84}
 P.M.Stevenson,
  \tarttitle{Gaussian Effective Potential: Quantum Mechanics}
 Phys.Rev.D \vol{30} (1984) 1712.

\bibitem{SSS}
A.N.Sissakian, I.L.Solovtsov and O.P.Solovtsova, Phys.Lett.B
\vol{321} (1994) 381.

\bibitem{Kl93}
 H.Kleinert,
 Phys.Lett.\vol{A173} (1993) 332.

\bibitem{JK95}
 W.Janke and H.Kleinert,
 Phys.Rev.Lett.\vol{75} (1995) 2787.


\bibitem{KSF}
 H.Kleinert and V.Schulte-Frohlinde,
 \tbktitle{Critical Properties of $\phi^4$-Theories}
 (World Scientific, Singapore, 2001).


\bibitem{KPP97}
 F.\ Karsch, A.\ Patk\'os and P.\ Petreczky,
 \tarttitle{Screened Perturbation Theory}
 Phys.Lett.B \vol{401} \yr{1997} 69
 [\eprint{hep-ph/9702376}].


\bibitem{Yu}
V.I.Yukalov, J.Math.Phys. \vol{33} (1992) 3994.


\bibitem{WZZSDYX}
C.M.Wu et al.,
 \tarttitle{Phase Structure of Lattice $\phi^4$
Theory by Variational Cumulant Expansion}
 Phys.Lett.B \vol{216} (1989) 381.
 \tcomment{\\ Lattice real phi**4 M1FE}

\bibitem{KM} W.\ Kerler and T.\ Metz,
\tarttitle{High-order effects in
  action-variational approaches to lattice gauge theory}
  Phys.Rev.D \vol{44} (1991) 1263.
 \tcomment{\\ (Pure gauge SU(2), FAC)}


\bibitem{BDJ}
 I.R.C.\ Buckley, A.\ Duncan and H.F.\ Jones,
 \tarttitle{Proof of the convergence of the linear $\delta$ expansion: Zero
  dimensions}
  Phys.Rev.D \vol{47} (1993) 2554.
 \tcomment{\\ (0 Dim. convergence of Z)}

\bibitem{BeDJ}
 C.M.Bender, A.Duncan and H.F.Jones,
 Phys.Rev.D \vol{49} (1994) 4219.
 \tcomment{\\ 0 Dim. convergence of ln(Z)}

\bibitem{Ok87}
 A.Okopi\'{n}ska,
 \tarttitle{Nonstandard expansion techniques for the effective potential in
      $\phi^4(x)$ quantum field theory}
      Phys.Rev\vol{D35} (1987) 1835.


\bibitem{PO98}
 S.A.\ Pernice and G.\ Oleaga,
 \tarttitle{Divergence of
  perturbation theory: Steps towards a convergent series}
  Phys.\ Rev.\ D \textbf{57} (1998) 1144\tpre{
  [\eprint{hep-th/9609139}]}.


\bibitem{BMPS}
C.M.Bender, K.A.Milton, S.S.Pinsky and L.M.Simmons Jr.,
J.Math.Phys.\ \vol{30} (1989) 1447.
 \tcomment{\\  Non Linear delta expansion, ODEs}

\bibitem{Ca}
W.E.Caswell, Ann.Phys. \vol{123} (1979) 153.
 \tcomment{\\  QM, PMS}

\bibitem{Ki}
J.Killingbeck, J.Phys.A \vol{14} (1981) 1005.
 \tcomment{\\ QM}



\bibitem{DJ} A.Duncan and H.F.Jones,
 \tarttitle{Convergence proof for optimised $\delta$ expansion: Anharmonic oscillator}
  Phys.Rev.D \vol{47} (1993) 2560.
 \tcomment{\\  QM}

\bibitem{GKS}
 R.Guida, K.Konishi and H.Suzuki,
 Ann.Phys.\ \vol{249} (1996) 106.

\bibitem{Ok87b}
  A.\  Okopi\'nska,
  \tarttitle{Nonstandard expansion techniques for the
  finite-temperature effective potential in $\lambda \phi^4$ quantum
  field theory}
  Phys.\ Rev.\ D \textbf{36} (1987) 2415.

\bibitem{GM}
 S.K.Gandhi and A.J.McKane,
 \tarttitle{Critical Exponents in the $\delta$-expansion}
  Nucl.Phys.B \vol{419} (1994) 424

\bibitem{OS98a}
K.~Ogure and J.~Sato,
 \tarttitle{Critical exponents and critical amplitude ratio of the scalar model
from finite-temperature field theory}
 Phys.\ Rev.\ D {\bf 57}, 7460 (1998)
  [\eprint{arXiv:hep-ph/9801439}].


\bibitem{Chi00}
 S.Chiku,
 \tarttitle{Optimised perturbation Theory at Finite Temperature -
two-loop analysis}
 Prog.Th.Phys \vol{104} \yr{2000} 1129,
 \eprint{hep-ph/0012322}

\bibitem{Meu}
 Y.Meurice,
 \tarttitle{A Simple Method to Make Asymptotic Series of Feynman
 Diagrams Converge}
 Phys.Rev.Lett. 88 (2002) 141601\tpre{[ \eprint{hep-th/0103134}]}.

\bibitem{BR02a}
 E.Braaten and E.Radescu,
 \tarttitle{Convergence of the Linear $\delta$ expansion in the Critical $O(N)$ Field Theory}
 Phys.Rev.Lett. \vol{89} (2002) 271602\tpre{[\eprint{hep-ph/0206108}]}.

\bibitem{OS99}
K.~Ogure and J.~Sato,
 \tarttitle{Critical exponents of O(N) scalar model at temperatures below the  critical value
 using auxiliary mass
 method}
 Prog.\ Theor.\ Phys.\  {\bf 102}, 209 (1999)
 [\eprint{hep-ph/9905370}].


\bibitem{EJW} T.S.\ Evans, H.F.\ Jones and D.\ Winder,
  \tarttitle{Non-perturbative calculations of a global $U(2)$ theory at
  finite density and temperature}
  Nucl.\ Phys.\ B \textbf{598} (2001) 578.



\bibitem{AJ93a}
 J.O.Akeyo and H.F.Jones,
 Phys.Rev.D \vol{47} (1993) 1668.

\bibitem{AJ93b}
 J.O.Akeyo and H.F.Jones,
 Z.Phys.C \vol{58} (1993) 629.

\bibitem{AJP}
 J.O.Akeyo, H.F.Jones and C.S.Parker,
 \tarttitle{Extended Variational Approach to the SU(2) Mass Gap on the Lattice}
 Phys.Rev.D \vol{51} (1995) 1298\tpre{ [\eprint{hep-ph/9405311}]}.

\bibitem{ZTW}
 X.-T.Zheng, Z.G.Tan and J.Wang,
 Nucl.Phys.B \vol{287} (1987) 171.

\bibitem{Ya}
 J.M.Yang,
 J.Phys.G \vol{17} (1991) L143.

\bibitem{YWZ}
 J.M.Yang, C.M.Wu and P.Y.Zhao,
 J.Phys.G \vol{18} (1992) L1.

\bibitem{ZL}
 X.-T.Zheng, B.S.Liu,
 Intl.J.Mod.Phys.A \vol{6} (1991) 103.

\bibitem{TZ} C.-I.\ Tan and X.-T.\ Zheng,
 \tarttitle{Variational-cumulant
  expansion in lattice gauge theory at finite temperature}
  Phys.\ Rev.\ D \vol{39} (1989) 623.



\bibitem{EIM}
 T.S.\ Evans, M.\ Ivin and M.\ M\"obius,
 \tarttitle{An optimised perturbation expansion for a global O(2) theory}
 Nucl.\ Phys.\ B \vol{577} (2000) 325.

\bibitem{Win} D.\ Winder,
 \tbktitle{Investigating Phase Transitions using the Linear Delta Expansion},
   Ph.D.\ Thesis, Imperial College,
  University of London, UK, 2001.

\bibitem{EJR98a}
 T.S.Evans, H.F.Jones and A.Ritz,
 \tarttitle{An Analytical
Approach to Lattice Gauge-Higgs Models} in \tinproctitle{Strong
and Electroweak Matter '97}, ed. F.Csikor and Z.Fodor (World
Scientific, Singapore, 1998\tpre{, ISBN {\tt
981-02-3257-8}})\tpre{ [\eprint{hep-ph/9707539}]}.

\bibitem{EJR98b}
 T.S.Evans, H.F.Jones and A.Ritz,
 \tarttitle{On the Phase Structure of the 3D SU(2)-Higgs Model and the Electroweak Phase
Transition}
 \journal{Nucl.Phys.} \vol{B517} (1998) 599\tpre{
[\eprint{hep-ph/9710271}]}.

\bibitem{HPV} M.\ Hasenbusch, K.\ Pinn and S.\ Vinti,
 \tarttitle{Critical Exponents of the 3D Ising Universality Class From Finite Size
  Scaling With Standard and Improved Actions}
  \verb|arXiv:hep-lat/9806012|.

\bibitem{BLH}
 H.W.J.\ Bl\"ote, E.\ Luijten and J.R.\ Heringa,
  \tarttitle{Ising universality in three dimensions: a Monte Carlo study}
  J.\ Phys.\ A \textbf{28} (1995) 6289.



\bibitem{Chang}
 S.-J.Chang,
 \tarttitle{Quantum fluctuations in a $\phi^4$
  theory. I. Stability of the vacuum}
  Phys.\ Rev.\ D \textbf{12} (1975) 1071.

\bibitem{HS}
 I.G.Halliday and P.Suranyi,
 \tarttitle{Anharmonic oscillator: A new approach}
 Phys.\ Rev.\ D \textbf{21} (1980) 1529.


\bibitem{Wo}
M.Wortis, \tarttitle{Linked Cluster Expansion} in
\tinproctitle{Phase Transitions and Critical Phenomena}, vol.3,
eds. C.Domb and M.S.Green (Academic Press, London, 1974).
 \tcomment{\\ Linked Cluster Expansion.}

\bibitem{Englert63}
 F.Englert,
 \tarttitle{Linked Cluster Expansions in
  the Statistical Theory of Ferromagnetism}
  Phys.\ Rev.\ \textbf{129} (1963) 567.


\bibitem{Rushbrooke64}
 G.S.\ Rushbrooke,
 \tarttitle{On the Theory of
  Randmly Dilute Ising and Heisenberg Ferromagnets}
  J.\ Math.\ Phys.\ \textbf{5} (1964) 1106.

\bibitem{RBW} G.S.\ Rushbrooke, G.A.\ Baker, Jr.\ and P.J.\ Wood, in
  \tbktitle{Phase Transitions and Critical Phenomena}, Volume 3, edited
  by C.\ Domb and M.S.\ Green
  (Academic Press, 1974).

\bibitem{HR99}
 H.Meyer-Ortmanns and T.Reisz,
 Nucl.Phys.B (Proc.Suppl.) \vol{73} (1999) 892\tpre{ [\eprint{hep-th/9809107}]}.


\bibitem{CHPPV}
 M.Campostrini et al.,
 \tarttitle{Critical behaviour of the three-dimensional XY universality class}
 Phys.Rev.B \textbf{63} (2001) 214503.

\bibitem{Bollobas}
 B.\ Bollob\'as,
 \tbktitle{Modern Graph Theory}
 (Springer, 1998).

\bibitem{Brout1}
 R.\ Brout,
 \tarttitle{Statistical Mechanical Theory of a
  Random Ferromagnetic System}
   Phys.\ Rev.\ \textbf{115} (1959) 824;

\bibitem{Brout2}
  R.Brout,
  \tarttitle{Statistical Mechanical Theory of
  Ferromagnetism. High Density Behavior}
   Phys.\ Rev.\ \vol{118}   (1960) 1009;

\bibitem{Brout3}
  R.Brout,
  \tarttitle{Statistical Mechanics of
  Ferromagnetism; Spherical Model as High-Density Limit}
  Phys.\ Rev.\ \vol{122} (1961) 469.

\bibitem{HC}
 G.\ Horwitz and H.B.\ Callen,
 \tarttitle{Diagrammatic
  Expansion for the Ising Model with Arbitrary Spin and Range of
  Interaction}
  Phys.\ Rev.\ \vol{124} (1961) 1757.

\bibitem{SCH}
  B.\ Strieb, H.B.\ Callen and G.\ Horwitz,
  \tarttitle{Cluster Expansion for the
  Heisenberg Ferromagnet}
  Phys.\ Rev.\ \vol{130} (1963) 1798.



\bibitem{Sherman64}
 S.\ Sherman,
 \tarttitle{Product Property and Cluster
  Property Equivalence}
  J.\ Math.\ Phys.\ \vol{5} (1964) 1137.



\bibitem{SA}
 D.Stauffer and A.Aharony,
 \tbktitle{Introduction to Percolation Theory}
 (Taylor and Francis, 1994).

\bibitem{Me90}
 S.Mertens,
 \tarttitle{Lattice Animals: A Fast Enumeration Algorithm and New
Perimeter Polynomials}
 J.Stat.Phys. \vol{58} (1990)  1095

\bibitem{nauty}
 B.D.McKay,
 \emph{nauty},
 \verb|http://cs.anu.edu.au/~bdm/nauty/|

\bibitem{Maple}
 Waterloo Maple,
 \verb|http://www.maplesoft.com/|

\bibitem{NR}
 W.H.Press, S.A.Teukolsky, W.T.Wetterling and
  B.P.Flannery,
  \tbktitle{Numerical Recipes in C++},
  (Cambridge University Press, Second Edition,2002).

\bibitem{GMP} GNU Multiple Precision Library,
  \verb|http://www.swox.com/gmp|

\bibitem{FL}
 A.M.Ferrenberg and D.P.Landau,
 \tarttitle{Critical behavior of the three-dimensional Ising model:
A high-resolution Monte Carlo study}
 Phys.Rev.B\vol{44} (1991) 5081.


\bibitem{TB} A.L.\ Talapov and H.W.J.\ Bl\"ote,
 \tarttitle{The
  magnetization of the 3D Ising model}
  J.\ Phys.\ A \vol{29}   (1996) 5727.


\bibitem{PDG}
  K. Hagiwara et al.,
  \tarttitle{The Review of Particle Physics}
  Physical Review D\vol{66}, 010001 (2002)




\bibitem{HMP}
 K.\ Huang, E.\ Manousakis and J.\ Polonyi,
 \tarttitle{Effective potential in scalar field theory},
 Phys.\ Rev.\ D \vol{35} (1987) 3187.

\bibitem{KS}
 J.\ Kuti and Y.\ Shen,
 \tarttitle{Supercomputing the Effective Action}
 Phys.\ Rev.\ Lett.\ \vol{60} (1988) 85.


\bibitem{ID}
 C.Itzykson and J.-M.Drouffe,
 \tbktitle{Statistical Field Theory, vol 2}
 (Cambridge University Press, Cambridge, 1989).




\bibitem{Mathematica} Mathematica, \verb|http://www.wolfram.com/|

\bibitem{Chiku00}
 S.\ Chiku,
 \tarttitle{Optimised Perturbation Theory at
  Finite Temperature --- Two-Loop Analysis ---}
  Prog.\ Theor.\ Phys.\ \vol{104} (2000) 1129.































\end{thebibliography}
\end{document}